\documentclass[]{aa}
\usepackage{natbib,twoopt}
\usepackage{amsmath}

\usepackage{color}
\usepackage[breaklinks=true]{hyperref} 
\bibpunct{(}{)}{;}{a}{}{,}             
\makeatletter
  \newcommandtwoopt{\citeads}[3][][]{\href{http://adsabs.harvard.edu/abs/#3}%
    {\def\hyper@linkstart##1##2{}%
     \let\hyper@linkend\@empty\citealp[#1][#2]{#3}}}
  \newcommandtwoopt{\citepads}[3][][]{\href{http://adsabs.harvard.edu/abs/#3}%
    {\def\hyper@linkstart##1##2{}%
     \let\hyper@linkend\@empty\citep[#1][#2]{#3}}}
  \newcommandtwoopt{\citetads}[3][][]{\href{http://adsabs.harvard.edu/abs/#3}%
    {\def\hyper@linkstart##1##2{}%
     \let\hyper@linkend\@empty\citet[#1][#2]{#3}}}
  \newcommandtwoopt{\citeyearads}[3][][]%
    {\href{http://adsabs.harvard.edu/abs/#3}
    {\def\hyper@linkstart##1##2{}%
     \let\hyper@linkend\@empty\citeyear[#1][#2]{#3}}}
\makeatother

\usepackage{graphicx}
\usepackage{txfonts}
\begin{document}

   \title{Probing star formation and ISM properties using galaxy disc inclination II}
   \subtitle{Testing typical FUV attenuation corrections out to z$\sim$0.7}
   
\authorrunning{S. K. Leslie et al.}

   \author{S. K. Leslie
          \inst{1}\thanks{Fellow of the International Max Planck Research School 
for Astronomy and Cosmic Physics at the University of Heidelberg (IMPRS-HD)}
          \and E. Schinnerer \inst{1}
                  \and B. Groves \inst{2}
         \and M. T. Sargent \inst{3}
          \and G. Zamorani \inst{4}
          \and P. Lang \inst{1}
         \and E. Vardoulaki\inst{5}
          }

   \institute{Max-Planck-Institut f\"{u}r Astronomie, K\"{o}nigstuhl 17, 69117, 
Heidelberg, Germany
\and Research School of Astronomy and Astrophysics, Australian National University, Canberra, ACT 2611, Australia
\and Astronomy Centre, Department of Physics and Astronomy, University of Sussex, Brighton BN1 9QH, UK
\and INAF-Osservatorio Astronomico di Bologna, via Gobetti 93/3, 40129, Bologna, Italy
\and Argelander-Institut f\"{u}r Astronomie, Auf dem H\"{u}gel 71, D-53121 Bonn, Germany\\
\email{leslie@mpia.de}
             }

   \date{Version 12 September 2017}

\abstract{ 
We evaluate dust-corrected far-ultraviolet (FUV) star formation rates (SFRs) for samples of star-forming galaxies at $z\sim0$ and $z\sim0.7$ and find significant differences between values obtained through corrections based on UV colour, from a hybrid mid-infrared (MIR) plus FUV relation, and from a radiative transfer based attenuation correction method. The performances of the attenuation correction methods are assessed by their ability to remove the dependency of the corrected SFR on inclination, as well as returning, on average, the expected population mean SFR. We find that combining MIR (rest-frame $\sim$13$\mu$m) and FUV luminosities gives the most inclination-independent SFRs and reduces the intrinsic SFR scatter of the methods we tested. However, applying the radiative transfer based method gives corrections to the FUV SFR that are inclination independent and in agreement with the expected SFRs at both $z\sim0$ and $z\sim0.7$. SFR corrections based on the UV-slope perform worse than the other two methods we tested. For our local sample, the UV-slope method works on average, but does not remove inclination biases. At z$\sim$0.7, we find that the UV-slope correction we used locally flattens the inclination dependence compared to the raw FUV measurements, but was not sufficient to correct for the large attenuation observed at z$\sim$0.7.}

   \maketitle
%

\section{Introduction}
Robust measurements of star formation rates (SFR) are a key ingredient in the context of galaxy evolution studies.
The far-ultraviolet (FUV; $\sim$1550\AA~) is dominated by radiation from O and B stars and thus traces recent star formation, provided the timescale for significant fluctuations is longer than a few 10$^7$ years \citep{Madau2014}). Extragalactic UV SFR calibrations have been revolutionised by the launch of the Galaxy Evolution Explorer mission (GALEX; \citealt{Martin2005}), which imaged approximately two-thirds of the sky at FUV and near-UV (NUV) wavelengths. 
\cite{Murphy2011} and \cite{Hao2011} used the Starburst99 stellar population synthesis code \citep{Leitherer1999} to derive conversions between GALEX FUV luminosity and SFR at an age of 100 Myr, assuming a constant SFR and solar metallicity:

\begin{equation}\label{EQsfruv}
\log\left(\frac{\text{SFR}}{\text{M}_\odot \text{ yr}^{-1}} \right)=\log\left(\frac{L_\text{FUV}}{\text{erg s}^{-1}}\right)-43.35.
\end{equation}

The great disadvantage for UV-based SFR measurements is that the light is heavily attenuated by dust. 
When stars are not individually resolved, the effective reddening of the emerging radiation is influenced by the dust-stars-gas geometry of the observed region \citep{Natta1984, Calzetti1997}. Furthermore, the reddening laws are not the same for all galaxies, nor for regions within galaxies, due to a combination of different dust grain size distributions and compositions, and differing geometries. 
The amount of unattenuated starlight is negligible in dusty starbursts, but is almost 100\% in dust-poor dwarf galaxies.

Dust is present in stellar birth clouds and in the diffuse interstellar medium, and both components absorb starlight and re-emit it in the infrared (IR). However, the dust--star geometry appears to evolve with redshift, with results from paper I in this series (\citealt[L18]{Leslie2017}) indicating that the fraction of dust in stellar birth clouds was higher at earlier times (z$\sim$0.7).

To infer dust-unbiased, total SFRs, corrections accounting for UV attenuation are necessary. A promising way to obtain the total SFR of a galaxy is to combine the UV emission with infrared emission, thereby capturing both the attenuated and unattenuated emission. Combined UV+IR SFRs have been used with success out to at least $z<3$ \citep{Wuyts2011a, Santini2014, Magnelli2014}.
However, older stellar populations can contribute to dust heating and infrared emission \citep{Bendo2010, Groves2012}. Therefore, the exact relation between infrared emission and recent SFR will depend on the relative contribution of the young and old stellar populations to the general stellar radiation field, which varies from galaxy to galaxy, see for example \cite{Boquien2016}.

Corrections at high redshift ($z>2$) often rely on empirical calibrations based on the UV spectral slope or UV colours.  
However, these calibrations are uncertain due to potential intrinsic variations in the UV slope and dust attenuation \citep{Battisti2016, Salmon2016}
Further testing of the UV slope and hybrid UV+IR attenuation correction methods are still required, particularly to ascertain that they are applicable at $z>0$.

We can evaluate UV-attenuation corrections by testing whether the corrected SFR is independent of galaxy inclination.
Inclined galaxies suffer from more line-of-sight attenuation than face-on galaxies because light has a larger column to pass through before escaping the galaxy. An ideal correction would account for this extra attenuation and result in a total SFR that is independent of viewing angle.
In L18,  we used the inclination - SFR (uncorrected for attenuation) relation of star-forming galaxies at different wavelengths (UV, mid-IR, far-IR, and radio) to probe the disc opacity of two matching samples of massive (M$_* >1.6\times 10^{10}$ M$_\odot$), large ($r_{1/2}>4$ kpc), disc-dominated ($n<1.2$) galaxies: a local Sloan Digital Sky Survey (SDSS) sample at z$\sim$0.07 and a sample of discs at z$\sim$0.7 drawn from the Cosmic Evolution Survey (COSMOS). We found that correcting the FUV flux for dust attenuation is critical even for face-on galaxies because the observed UV fluxes can underestimate the true SFR by a factor of $\sim$1.6 for face-on massive star-forming galaxies at z$\sim$0 and $\sim$5.5 at z$\sim$0.7.

In this paper we test the inclination dependence of UV-derived SFRs that have been corrected for attenuation using common empirical approaches: UV-slope corrections (Section \ref{uvslope}), hybrid UV+MIR SFRs (Section \ref{sechybrid}), and SFRs corrected for the attenuation expected from the radiative transfer model of \cite{Tuffs2004} and \cite{Popescu2011} (Section \ref{RT}). We apply these attenuation corrections to two samples of massive star-forming galaxies, $\sim$3500 galaxies at $z\sim0$ and $\sim$1000 galaxies at $\sim0.7$, and test for inclination dependence following the methodology of L18.

Throughout this work, we use the \cite{Kroupa2001} initial mass function (IMF) and assume a flat $\Lambda$ CDM (cold dark matter) cosmology 
with ($H_0$, $\Omega_M$, $\Omega_\Lambda$)= (70 km s$^{-1}$ Mpc$^{-1}$, 
0.3,0.7).
\section{Data and sample selection}
L18 described the data and sample selection, which we only briefly review here. We selected a local sample ($0.04<z<0.1$) from the spectroscopic SDSS matched with the GALEX Medium Imaging Survey (MIS) by \cite{Bianchi2011}. Our intermediate-redshift sample ($0.6<z<0.8$) from the COSMOS field is centred at $z\sim 0.7,$ where we have high-resolution rest-frame morphological measurements in the I band \citep{Sargent2007} that match our SDSS g-band morphological measurements \citep{Simard2011} in terms of rest-frame wavelength. Additionally, the GALEX NUV filter at z$\sim 0.7$ matches the FUV filter at z$\sim$0.1, ensuring that our FUV fluxes obtained from the GALEX Deep Imaging Survey (DIS) do not require k corrections.

L18 found that inclination affects the derived galaxy properties that are used for sample selection. After fitting and subtracting a power law from the relations of inclination versus galaxy property for $M_*$, g-band half-light radius $r_{1/2}$ , and S\'{e}rsic index $n$, we obtained ``face-on'' equivalent values for each galaxy. The inclination-corrected parameters, denoted with the subscript ``corr'', were only used for sample selection. 
Our selections from L18 are as follows and are referred to as the ``restricted'' sample:
\begin{itemize}
\item log(M$_{*,\textrm{corr}}$/M$_\odot$)$>$10.2,
\item $n_\textrm{corr}<$1.2,
\item r$_\textrm{1/2,corr}>$4 kpc
\item $L_\text{FUV}/\sigma_{L_\text{FUV}}>3$.
\end{itemize}
Star-forming galaxies were selected using their emission line ratios or colours for the local and COSMOS samples, respectively. The restricted sample contains 1021 galaxies in the local sample and 295 galaxies in the $z\sim0.7$ sample. 
The size cut ensures completeness in surface brightness, and the S\'{e}rsic index cut reduces contamination by bulge-dominated galaxies. However, to make our results applicable to a wider population of galaxies, we also present results for a sample that is not subject to size restrictions and has a less restrictive constraint on the S\'{e}rsic index:
\begin{itemize}
\item log(M$_{*,corr}$/M$_\odot$)$>$10.2,
\item $n_{corr}<$2
\item $L_\text{FUV}/\sigma_{L_\text{FUV}}>3$.
\end{itemize}
We refer to this sample as the ``full'' sample, and it contains 3489 galaxies at $z\sim$0 and 575 galaxies at $z\sim0.7$. 
Before requiring an FUV detection at a signal-to-noise ratio >3, the full local sample contains 8576 galaxies and the full COSMOS sample contains 1135 galaxies. Although the signal-to-noise ratio cut biases the sample towards galaxies with higher SFRs, it is necessary for reliable results. In our COSMOS sample we are likely missing the most highly inclined ($1-cos(i)>0.75$; $i=75.5^o$) galaxies with lower $SFR_\text{FUV}$ because of the sensitivity limit of GALEX DIS data.

In addition to the previous UV and MIR samples of L18, we also include here a UV-slope-corrected sample (section 3.1), which requires the addition of rest-frame NUV data. This results in a smaller subsample, especially at $z\sim0.7$ (Table 1).

\section{Corrected FUV star formation rates, SFR$_{\text{FUV,corr}}$}\label{uvcorr}

In this section, we apply three common SFR$_{\text{FUV}}$ attenuation corrections to test whether any residual trends of SFR with inclination persist.

\begin{figure*}
 \includegraphics[width=\linewidth]{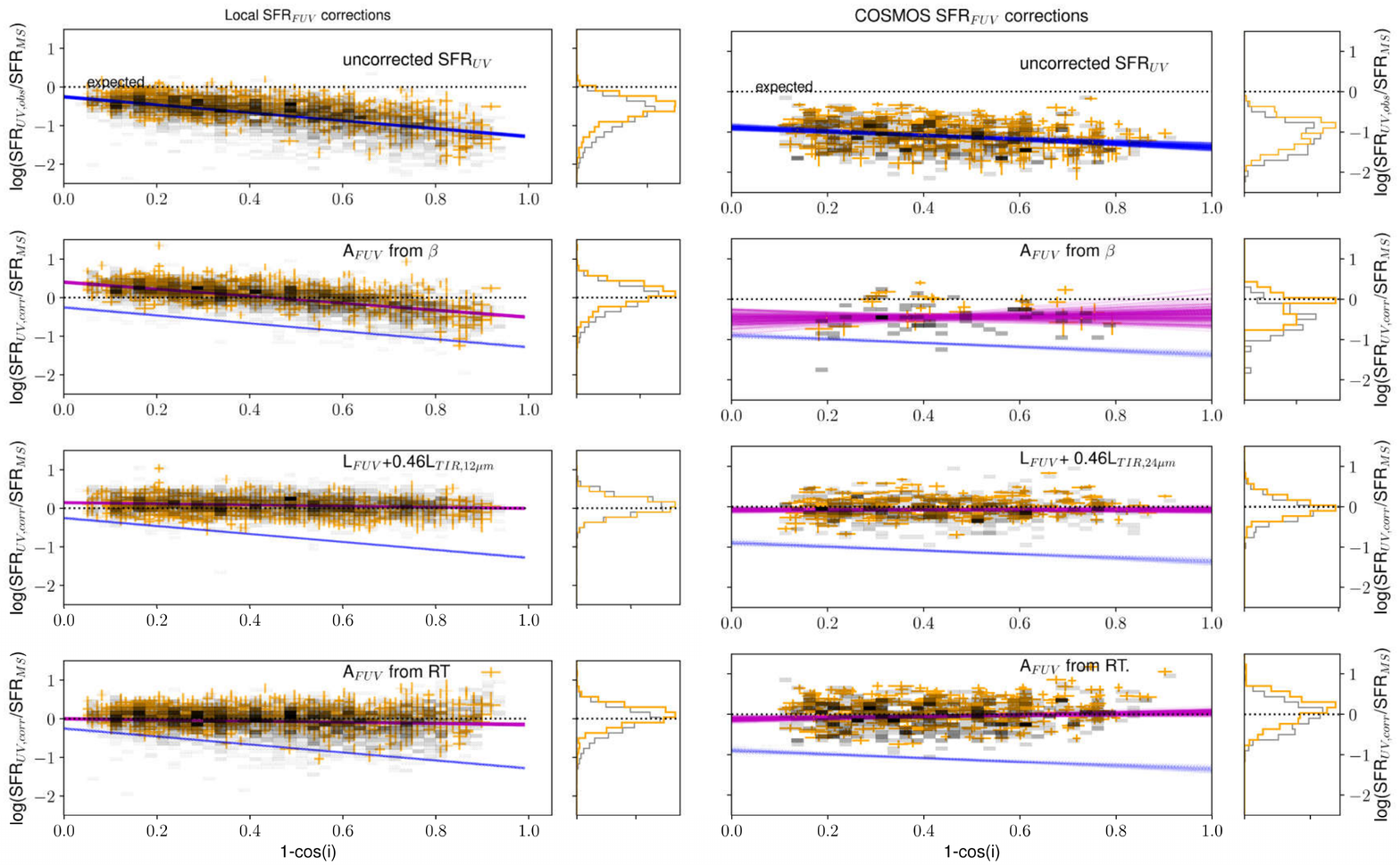}
\caption{Evaluating the inclination dependence of dust-corrected SFR$_\text{FUV}$ values. From top to bottom, the panels show the observed UV SFR, the SFR corrected based on the UV-slope, the hybrid MIR+UV determined SFR, and the SFR corrected based on the \cite{Tuffs2004} radiative transfer models as a function of inclination angle. Face-on galaxies are lie to the left, 1-$\cos(i) \sim0$, and edge-on galaxies lie to the right, 1-$\cos(i)\sim1$. The blue lines show the inclination relation for SFR$_{\text{FUV}}$ as observed with no dust corrections (top panel). The dust correction factors work to increase the SFR$_{\text{FUV}}$ and lessen the inclination dependence of the SFR. Fitted parameters are given in Table \ref{tablecorr}. Orange data points show the ``restricted
sample", and the full sample is shown as a 2D density histogram. Magenta lines show 150 random realisations of linear parameters from our MCMC chain. Distributions of the y-axes, log(SFR$_{\text{FUV}}$/SFR$_\text{MS}$) are also displayed as histograms. The dotted line shows where the SFR$_{\text{FUV}}$ = SFR$_\text{MS}$. }\label{fuvcorr}
\end{figure*}

We chose to normalize the SFR$_{\text{UV,corr}}$ by the SFR expected for a galaxy lying exactly on the galaxy SFR-stellar mass relation (galaxy main-sequence, MS).
As in L18, we adopted the best-fit MS relation for an updated version of the $z\lesssim1$ data compiled in
\cite{Sargent2014} (private communication),\begin{equation}\label{MSeq}
\log\left(\frac{\text{SFR}_\mathrm{MS}}{\text{M}_\odot\text{yr}^{-1}}\right) = 0.816 \log\left(\frac{M_*}{\text{M}_\odot}\right) - 8.248 + 3\log(1+z).
\end{equation}
We took into account the different IMF \citep{Chabrier2003} and cosmology (WMAP-7;
\citealt{Larson2011}) assumed by \cite{Sargent2014} (see L18 for more details).
Figure \ref{fuvcorr} shows the relation between log(SFR$_{\text{FUV,corr}}$/SFR$_\text{MS}$) and galaxy inclination (1-cos($i$)), where SFR$_\text{MS}$ is the main-sequence SFR expected given the mass and redshift of each galaxy.

The fits shown in Figure \ref{fuvcorr} are of the form \begin{equation}\log\left(\text{SFR}_\text{FUV,corr}/\text{SFR}_\text{MS}\right) = k_1(1-\cos(i)) + k_2.\end{equation} Best fits were obtained using the Monte Carlo Markov chain implementation \textit{emcee} (a \textsc{Python} package) \citep{ForemanMackey2013}, incorporating errors on the SFRs and inclination and fitting for an intrinsic variance term, $\sigma^2$, following \cite{DFM17}. For simplicity, the intrinsic scatter was assumed to be in the y-direction only (SFR). 
We used 64 walkers that created a chain of 6000 samples each for a total of 340 000 samples. The first 150 samples were discarded as an initialisation period (burn-in). The fitted parameters for the original uncorrected and corrected SFR relations are shown in Table \ref{tablecorr}, where we give the median, and the 5th and 95th percentile of the MCMC distribution.

The top panels of Figure \ref{fuvcorr} show the observed SFR$_{\text{FUV}}$. The restricted samples span the same range of inclinations as the larger full samples. However, the restricted samples tend to be skewed towards higher SFR$_{\text{FUV}}$ values, showing a lack of SFR$_\text{FUV,obs}$ at all inclinations. The restricted samples' galaxies are larger and more disc dominated than the full samples. 
These values indicate a mean attenuation factor of $A_{\text{FUV}}\sim 1.9$ mag ($\sim1.5$ mag) at $z\sim0$ and $A_{\text{FUV}}\sim2.9$ mag ($\sim2.4$ mag) at $z\sim0.7$ for the full (restricted) samples. 

The fitted slope of the SFR$_{\text{FUV}}$-inclination for the COSMOS (-0.44$\pm$0.15) and local (-0.9$\pm$0.06) restricted samples differ from the slopes found in L18 (-0.54$\pm0.09$ and -0.79$\pm0.04$ for the COSMOS and SDSS samples, respectively) because different fitting methods were adopted. The fitting algorithm in this work is more lenient towards outliers, but was selected in order to evaluate the scatter of our SFR values before and after corrections. We chose to test the correction methods described below, which
are representative of those commonly applied in the literature. 

\subsection{UV slope correction}\label{uvslope}
The observed relationship between the rest-frame UV continuum slope $\beta$, where $F_\lambda \propto \lambda^\beta$, and the ratio of UV to IR light of a galaxy (IRX, log(LIR/LUV)) has been used to calibrate $\beta$ as an attenuation correction tool since the correlation was first observed for starburst galaxies by \cite{Meurer1999} and \cite{Calzetti1997}. 
Longer-lived ($\sim$200 Myr), lower-mass stars contribute substantially to the NUV emission at later times. For a constant SFR, the UV spectral slope therefore reddens moderately with time, complicating the conversion from luminosity into SFR. 
For this analysis we estimated $\beta$ by fitting $\log(F_\lambda) = \beta \log( \lambda (1+z) ) + c$ using the \textsc{Python} package \textsc{polyfit}, with data weighted by their flux errors. For our SFR$_{\text{FUV,corr}}$ error values, we used and propagated the errors on the fitted slope given by \textsc{polyfit}. When only two data points were available, we calculated the gradient and the corresponding error. 
We then used the \cite{Boquien2012} relation (see their Section 4.2),
\begin{equation}
A_\text{FUV} = 0.87(\beta +2.586),\label{beta}
\end{equation} to determine the amount of attenuation and correct SFR$_\text{FUV}$. 
We were unable to avoid photometric bands with emission or absorption lines, as suggested for example by \cite{Calzetti1994}, which increases the uncertainty $\beta$. However, \cite{Leitherer2011} found a good agreement between UV spectral slopes determined from International Ultraviolet Explorer spectra and from the GALEX FUV and NUV magnitudes of nearby galaxies. The $A_\text{FUV}$ -- $\beta$ relation in Eq. \ref{beta} was derived for normal face-on star-forming spiral galaxies. This relation varies from that of starburst galaxies (e.g. \citealt{Overzier2011,Meurer1999}), largely due to intrinsic UV colour differences between galaxies and yields systematically lower attenuation for a given $\beta$. 
Observations of a wide range of galaxies have shown a large scatter in the UV colour-extinction relation \citep{Boquien2012} and about two orders of magnitudes of scatter in actual FUV attenuations for a fixed FUV-NUV colour. Studies such as \cite{CatalanTorrecilla2015} have shown that even with a UV colour correction, the FUV alone is not a reliable tracer of SFR in individual galaxies. This is because of the large extinction correction needed at FUV wavelengths, and the variation of different attenuation laws in the UV.

\begin{figure}
\includegraphics[width = \linewidth]{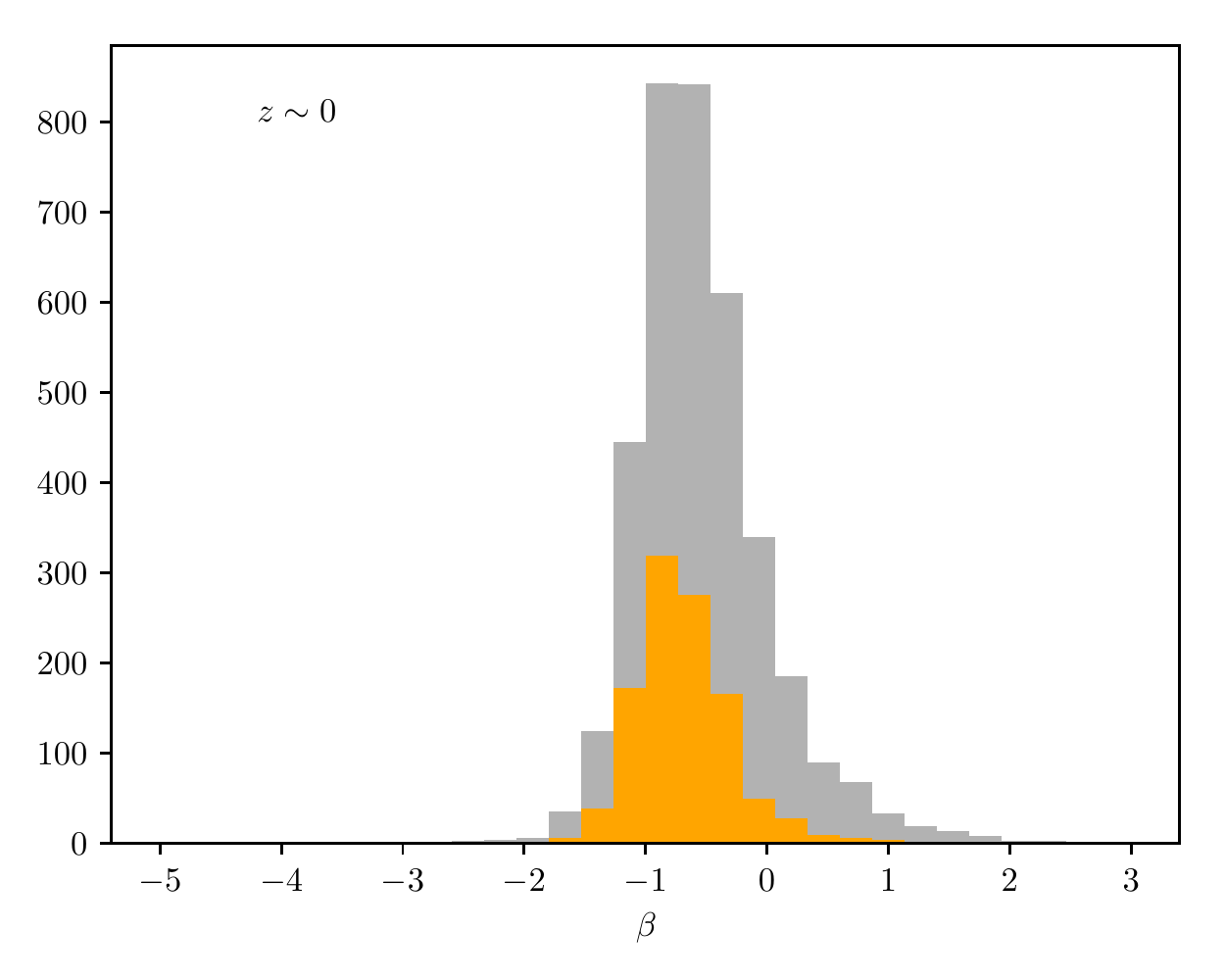}
\includegraphics[width = \linewidth]{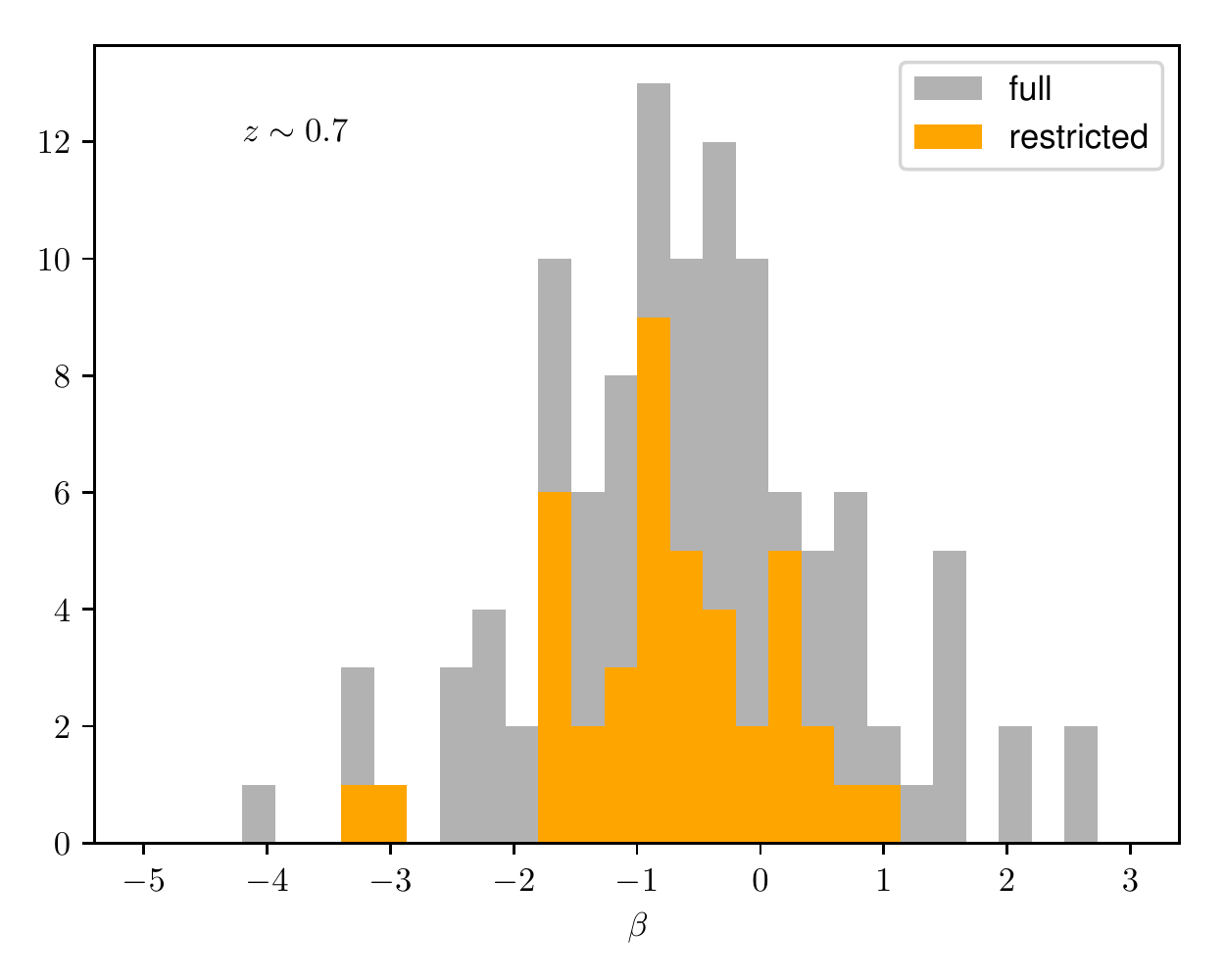}
\caption{UV-slope, $\beta$, distribution for restricted (orange) and full samples (grey) in our local sample (top) and $z\sim0.7$ sample (bottom).}
\label{betahist}\end{figure}

\paragraph{Local sample:}
For our SDSS sample, we have access to photometry from only two UV bands: GALEX NUV ($\lambda_{\text{eff}}$= 2274\AA~) and FUV ($\lambda_{\text{eff}}$= 1542\AA~). 
Using A$_\text{FUV}$ , we corrected the observed L$_{\text{FUV}}$ used to calculate SFR in Equation \ref{EQsfruv} and show the trend of SFR$_{\text{FUV,corr}}$ with the inclination in Figure \ref{fuvcorr}. Figure \ref{betahist} shows the distribution of $\beta$, peaking close to $\beta\sim-1$ for both the full and restricted samples.

The colour-corrected SFR$_{\text{FUV}}$ show a slightly shallower inclination dependency (slope of -0.8$\pm$0.02 instead of -1.0$\pm$0.03 for the full sample, and -0.79$\pm0.03$ instead of -0.9$\pm0.04$ for the restricted sample) and overall SFRs that are closer to the MS, but further improvements are required because the slopes are significantly different from zero and one will tend to over-estimate the SFR for face-on galaxies and under-estimate the SFR for edge-on galaxies. 


\paragraph{COSMOS sample:}
For the COSMOS sample, we fit the UV spectral slope $\beta$ using flux densities observed between rest-frame 1250-2500\AA. For our redshift range, this corresponds to the COSMOS2015 B, IB427, u, or GALEX NUV bands. 
To fit the spectral slope, we did not use flux values with a flag $>$0 \citep{Laigle2016}. Our data show $\beta$ values ranging between 2 and -3, with most galaxies having $\beta\sim-0.8$ for both the full and restricted samples. 

The limiting factor is the small number of photometric data points in the rest-frame wavelength range 1250 - 2500\AA~ over our redshift range. For $z<0.705$, only the NUV and $u$ band lie in the rest-frame 1250-2500\AA~ range. Between $0.70<z<0.78,$ the IB427 band can also be used to constrain $\beta,$ and for $0.78<z<0.8,$ we can use the u, IB247, and the B band along with GALEX NUV to constrain $\beta$. No redshift dependency on our fitted $\beta$ slopes arises because different bands were used for the full and restricted samples.

The SFR$_\text{FUV}$ corrected for attenuation using the spectral slope appears to remove most of the inclination dependence at z$\sim$0.7. The slopes are 0.05$\pm$0.45 for the full sample and -0.05$\pm$0.51 for the restricted sample; however, these slopes are poorly constrained because of the small number of galaxies (56 in the full and 28 in the restricted samples, respectively) with robust $\beta$ measurements at $z\sim 0.7$. The average SFR has also increased, but the corrected values are still below the MS by -0.46$\pm0.20$ dex for the full sample and -0.12$\pm0.25$ dex for the restricted sample for face-on galaxies. Figure \ref{betahist} shows the distribution of $\beta$ is very similar for the $z\sim0.7$ and $z\sim0$ samples, peaking around $\beta\sim -1$.

In Appendix \ref{A1} we explore the effects of assuming a constant attenuation factor of $A_V=1$ for all galaxies in the COSMOS sample. Assuming a constant factor is, in some cases, the only way to build a statistically significant sample of galaxies with SFR measurements. While this assumption ignores any inclination effects, it can correct the systematic offset observed in the top right panel of Figure 1. We also show the results for an alternative A$_\text{FUV}$--$\beta$ relation from \cite{Wang2016} in Appendix \ref{A1}.

\subsection{MIR hybrid correction}\label{sechybrid}
The most widely used composite tracers combine the UV observations with IR measurements using an energy-balancing approach that in the simple case assumes that the IR and UV light comes from a similar optical depth. This is generally a valid assumption and means that a linear combination of observed L$_{\text{FUV}}$ and L$_\text{IR}$ luminosity can correct the UV fluxes for dust attenuation \citep{Kennicutt2009, Hao2011, CatalanTorrecilla2015},
\begin{equation}
L_\text{UV,corr}=L_\text{UV,observed}+\eta L_\text{IR}
.\end{equation}
The coefficient $\eta$ can be calibrated theoretically with evolutionary synthesis models or empirically using independent measurements of dust-corrected SFRs \citep{Treyer2010, Hao2011}.
In most cases, $\eta<1$ because dust is significantly heated by old stars and by emission arising from the IR cirrus. Different sample selections result in different determinations of $\eta$ because of different contributions of the old and young stellar populations in the dust-heating, inter-stellar radiation field.
For example, early studies were restricted to luminous starburst galaxies and star-forming regions (e.g. \citealt{Buat1999, Meurer1999}), and yielded values of $\sim$0.6 \citep{Kennicutt2012}. With the advent of GALEX, \textit{Spitzer,} and \textit{Herschel}, subsequent analyses of normal star-forming galaxies typically find lower $\eta$ values, for instance \cite{Hao2011}, and \cite{Wang2016} (although see \citealt{Boquien2016}).
In this analysis, we have access to MIR fluxes through WISE (local sample) and Spitzer-MIPS (COSMOS sample). The MIPS 24$\mu$m data samples rest-frame 14$\mu$m fluxes at $z\sim0.7$. The short-wavelength MIR range ($\sim$3-20 $\mu$m) dust emission arises from strong polycyclic aromatic hydrocarbon (PAH) emission features and continuum \citep{Leger1984, Allamandola1985, Dopita2003, Smith2007}. The continuum emission comes from both single-photon, stochastically
heated small dust grains (typically $<$ 0.01$\mu$m) and from thermally emitting hot (T$>$150 K) dust \citep{Calzetti2013}.

We correct the UV luminosity following \cite{Hao2011}
\begin{equation}
L_{\text{FUV,corr}} = L_{\text{FUV,obs}}+0.46 L_{\text{IR}}.\label{hybrid}
\end{equation}
For our analysis we adopted SFR$_{\text{FUV,corr}}$ errors given by the uncertainties on the UV and MIR fluxes alone.
However, we note that $\eta$ has been found to vary with stellar mass and morphological type by up to factors of 4 for local late-type galaxies \citep{Boquien2016,CatalanTorrecilla2015}. 

\paragraph{Local sample:}
The WISE W3 band is broad, covering 7.5-16.5$\mu$m at half-power \citep{Jarrett2011}. This wavelength range encompasses multiple PAH features as well as H$_2$ and forbidden neon emission lines. PAH fractions tend to be high in regions of active star formation, and the WISE 12$\mu$m emission has been shown to correlate well with SFR \citep{Cluver2014, Cluver2017}. 
We followed the method in L18 to convert the W3 flux into total infrared luminosity using the spectral energy distribution (SED) template of \citealt{Wuyts2008}. Additionally, we discuss the use of other monochromatic-to-total infrared (TIR) calibrations making use of WISE 12$\mu$m and 22$\mu$m data in Appendix \ref{A2}.

We then corrected our L$_{\text{FUV}}$ according to Equation \ref{hybrid} and calculated SFR$_{\text{FUV,corr}}$. The resulting SFR$_{\text{FUV,corr}}$ values show very little inclination dependence on slopes $k_1=-0.15\pm0.03$ and $-0.21\pm0.05$ for the full and restricted samples, respectively. The fitted intrinsic scatter of the hybrid SFR-inclination relation is smaller than the other local SFR-inclination relations. The corrected SFRs of face-on galaxies lie above the expected SFR$_{MS}$ by 0.15$\pm$0.02 and 0.14$\pm$0.02 dex for the edge-on and face-on samples, while the corrected SFRs of edge-on galaxies are entirely consistent with the MS. 

\paragraph{COSMOS sample:}
For the COSMOS sample, we have 24$\mu$m flux measurements, or rest-frame $\sim$14$\mu$m. We followed the method in L18 to convert this flux into total infrared luminosity (using the SED template of \citealt{Wuyts2008}). The resulting hybrid SFR$_{\text{FUV,corr}}$ --inclination relation has a slope consistent with zero for both the full and restricted subsamples. 
The y-intercept values are also low: -0.16$\pm$0.09 and -0.03$\pm$0.06 for the full and restricted samples, respectively. For the COSMOS hybrid corrections, the restricted sample has, on average, slightly higher SFR$_\text{FUV,corr}$ than the full sample. 
The scatter is not reduced as significantly as it was in the local sample for the hybrid SFRs, but is still the smallest out of the four SFR$_\text{FUV}$--inclination relations shown. Other authors such as \cite{Izotova2018} also found a smaller scatter in the MIR+UV hybrid SFRs than for SFRs derived from either of the original UV or MIR bands.

\subsection{Radiative transfer correction}\label{RT}

Integrated SED models fitted to observations can constrain physical properties of a galaxy, including the SFR. The radiative transfer (RT) method self-consistently calculates the dust emission SED based on the radiation field of the attenuated stellar population. When creating model SEDs using the RT method, the geometry and distribution of the stars, dust, and gas in the galaxy can be determined either from a hydrodynamical model or a toy model of the galaxy. In L18, SFR$_{\text{FUV,obs}}$-inclination relations were fit with the model predictions of the \cite{Tuffs2004} RT calculations based on a simple geometric model for a spiral galaxy. For the UV attenuation calculations, it was assumed that UV emission comes only from the thin disc of a galaxy, simplifying the calculation by removing the bulge or thick-disc stellar model components. Under this assumption, the parameters required for the model are inclination angle,  B-band opacity through the centre of the galaxy as seen face-on ($\tau_B^f$), and the fraction attenuation occurring locally in a clumpy dust component associated with star-forming regions in the thin disc (as opposed to the diffuse dust component), $F$. 

The model of \cite{Tuffs2004} derives the attenuated stellar population, and the resulting dust emission expected was calculated in \cite{Popescu2011}. Recently, \cite{Grootes2013}, \cite{Davies2016}, and \cite{Grootes2017}, have used RT dust corrections based on the \cite{Tuffs2004} and \cite{Popescu2011} models to correct NUV luminosities for dust attenuation.

\cite{Tuffs2004} fit the attenuation curves (the change of magnitude $\Delta m$ due to dust attenuation; $\Delta m$ vs 1-$\cos(i)$) of the thin-disc component with a fifth-order polynomial at various $\tau_B^f$ values and wavelengths. We interpolated the \cite{Tuffs2004} models to derive the $\Delta m$ at 1542\AA, which is the effective wavelength of the GALEX FUV filter. Then $L_\text{FUV}$ was corrected with $\Delta m$ to give SFR$_\text{FUV,corr}$. 

For the errors on the SFRs corrected by this method, we adopted a constant $\sigma = 0.09$ dex as suggested by \cite{Davies2016} in their Appendix A. 
\cite{Davies2016} found that the RT method produces the most consistent slopes and normalisations of the SFR-M$_*$ relation out of the 12 different SFR metrics compared for local spiral galaxies\footnote{\cite{Davies2016} calculated $\tau_B^f$ for each galaxy by assuming an empirical relation between $\tau_B^f$ and stellar mass surface density \citep{Grootes2013}. We chose not to follow this assumption for our analysis as it requires a fixed $F=0.41$ and was not derived for galaxies at z$\sim$0.7}. 

\paragraph{Local sample:} 
We repeated our analysis of L18, Section 3.3, and fit the \cite{Tuffs2004} models to our full sample. We find best-fitting parameters of $\tau_B^f = 4.42^{+0.11}_{-0.11}$, $F=0.209^{+0.009}_{-0.01}$. We then used these fixed parameters to correct the SFR for each galaxy, given its inclination angle. 
Our best-fit slope values become significantly flatter than for the uncorrected sample, being -0.01$\pm$0.02 and -0.06$\pm$0.06 for the full and restricted samples, respectively. However, this correction does not reduce the observed dispersion $\sigma^2$ for the full sample. By using the best-fitting parameters for the full galaxy sample to correct $SFR_\text{FUV}$, we would expect the normalisation and intercept to be close to zero for the full sample. 

\cite{Grootes2017} used an empirical relation between stellar mass surface density and B-band optical depth \citep{Grootes2013} to set $\tau_B^f$, fixing $F=0.41$ \citep{Popescu2011}. Adopting this method gives ($k_1k_2,\sigma^2$)= (0.14$\pm$0.03, 0.18$\pm$0.02, 0.25$\pm$0.01) for the full and ($0.20\pm0.05, 0.22\pm0.02, 0.16\pm0.01$) for the restricted samples, respectively. This correction over-estimates the FUV SFRs on average, and leaves a positive residual inclination dependency. 
On the other hand, adopting the \cite{Grootes2013} $\tau_B^f$, and keeping our fitted value of $F=0.209$, we find ($k_1,k_2,\sigma^2$) = ($0.14\pm0.03,-0.03\pm0.02, 0.26\pm0.01$) for the full sample and ($0.19\pm0.05, 0.22\pm0.02,0.16\pm0.01$) for the restricted sample. Naturally, using different $\tau$ values for each galaxy reduces the scatter in the SFR$_\text{FUV,corr}$ values. However, the parameters of $\tau_B^f$ and $F$ are derived jointly in the best fit, and therefore must be set consistently with each other, as done for our adopted method in Table \ref{tablecorr} ($\tau_B^f = 4.42$, $F=0.209$ for all galaxies).


\paragraph{COSMOS sample:} 
We find best-fitting parameters for the full $z\sim0.7$ sample to be $\tau_B^f =2.85^{+0.24}_{-0.17}$ and $F=0.607^{+0.003}_{-0.006}$.
Using these values to determine the required UV attenuation correction, we obtain best-fit values that show a positive slope of 0.16$\pm$0.13 and 0.16$\pm$0.16 for the full and restricted samples, respectively. These slopes are different from zero at the 2 and 1.6 $\sigma$ levels for the full and restricted samples, respectively; moreover, it appears that they are driven by galaxies with the highest inclination angles, where the correction is most uncertain. The intrinsic scatter has increased from the scatter of the uncorrected SFRs by this correction.


\begin{table*}
\def\arraystretch{1.2}
\caption{Best-fit parameters of the relation $\log\left(SFR_{FUV,corr}/SFR_{MS}(M_*,z)\right) = k_1(1-cos(i)) + k_2$, where SFR$_{\text{FUV,corr}}$ is measured using three different methods. An intrinsic scatter of variance $\sigma^2$, assumed to be in the SFR direction, is also fitted following \cite{DFM17}. N is the number of galaxies used for the fits. Errors represent the 5th and 95th percentiles of the parameters given by the MCMC.}\label{tablecorr}
\centering
\begin{tabular}{|c|cccc|cccc|}
\hline
& \multicolumn{4}{c|}{Local $z\sim 0 $} & \multicolumn{4}{c|}{COSMOS $z\sim0.7$}\\
 Method & N & $k_1$ & $k_2$& $\sigma^2$ & N & $k_1$ & $k_2$ & $\sigma^2$\\
 \hline
No corr & 3489 & -1.03 $^{+0.04 }_{-0.04 }$ & -0.25 $^{+0.02 }_{-0.02 }$ & 0.31 $^{+0.01 }_{-0.01 }$& 
575&-0.48 $^{+0.12 }_{-0.12 }$ & -0.90 $^{+0.06 }_{-0.05 }$ & 0.28 $^{+0.02 }_{-0.02}$\\
UV slope & 3489& -0.91 $^{+0.03 }_{-0.03 }$ & 0.40 $^{+0.02 }_{-0.02 }$ & 0.25 $^{+0.01 }_{-0.01 }$& %
56 &0.05 $^{+0.45 }_{-0.44 }$ & -0.46 $^{+0.21 }_{-0.20 }$ & 0.29 $^{+0.06 }_{-0.07 }$\\
FUV + MIR & 3454&-0.15 $^{+0.03 }_{-0.03 }$ & 0.15 $^{+0.02 }_{-0.02 }$ & 0.20 $^{+0.01 }_{-0.01 }$& %
527 &0.02 $^{+0.15 }_{-0.15 }$ & -0.16 $^{+0.09 }_{-0.09 }$ & 0.27 $^{+0.02 }_{-0.02 }$\\
FUV + RT & 3489 & -0.15 $^{+0.04 }_{-0.04 }$ & -0.01 $^{+0.02 }_{-0.02 }$ & 0.31 $^{+0.02 }_{-0.02}$ & %
565 & 0.16 $^{+0.13 }_{-0.13 }$ & -0.12 $^{+0.06 }_{-0.06 }$ & 0.32 $^{+0.02}_{-0.02 }$\\
\hline
 & \multicolumn{4}{c|}{Restricted $z\sim0$ sample} & \multicolumn{4}{c|}{Restricted $z\sim0.7$ sample}\\
 \hline
No corr & 1021 &-0.90 $^{+0.06 }_{-0.06 }$ & -0.16 $^{+0.03 }_{-0.03 }$ & 0.23 $^{+0.01 }_{-0.01 }$&%
295  &-0.44 $^{+0.15 }_{-0.15 }$ & -0.75 $^{+0.07 }_{-0.07 }$ & 0.26 $^{+0.02 }_{-0.02}$\\
UV slope & 1021& -0.79 $^{+0.05 }_{-0.05 }$ & 0.45 $^{+0.03 }_{-0.03 }$ & 0.20 $^{+0.01 }_{-0.01 }$&%
28 &-0.05 $^{+0.51 }_{-0.50 }$ & -0.12 $^{+0.25 }_{-0.24 }$ & 0.25 $^{+0.07 }_{-0.09 }$\\
FUV + MIR & 1015 &-0.21 $^{+0.05 }_{-0.05 }$ & 0.14 $^{+0.02 }_{-0.02 }$ & 0.17 $^{+0.01 }_{-0.01 }$&%
284 & 0.10 $^{+0.12 }_{-0.12 }$ & -0.03 $^{+0.06 }_{-0.06 }$ & 0.22 $^{+0.02 }_{-0.02 }$\\
FUV + RT & 1021 &-0.06 $^{+0.06 }_{-0.06 }$ & 0.11 $^{+0.03 }_{-0.03 }$ & 0.21 $^{+0.01 }_{-0.01 }$& %
295 &0.16 $^{+0.16 }_{-0.16 }$ & 0.04 $^{+0.08 }_{-0.08 }$ & 0.31 $^{+0.02 }_{-0.02 }$\\

\hline
\end{tabular}
\end{table*}

\section{Discussion and conclusion}
We have tested a number of global galaxy FUV attenuation correction methods for samples of massive (M$_*>1.6\times10^{10}$) galaxies at $z\sim0$ and $z\sim0.7$ selected to be representative of corrections applied in the literature. 
FUV SFRs are very difficult to properly correct for dust attenuation.
Conversely, SFR calibrations at different wavelengths are sensitive to stellar populations over different timescales. SFR estimators at longer wavelengths are not able to capture variations on short timescales, leading to important discrepancies on the estimates of the instantaneous SFR. Despite these many uncertainties, we find that hybrid MIR and RT attenuation correction methods can remove the overall offset from the MS and decrease the inclination dependency of UV SFRs. 

Hybrid MIR+UV SFR calibrations remove a significant fraction of the inclination dependency displayed by uncorrected SFR$_\text{FUV}$ at z$\sim$0.7 and at z$\sim0$. However, there is some remaining uncertainty regarding the normalisation of the calibration due to differences in IR templates and intrinsic SFRs. The scatter of SFR from galaxy to galaxy is decreased when this technique is used. We also found that the RT method of \cite{Tuffs2004} works reasonably well for correcting the normalisation and inclination dependency of $\text{SFR}_{\text{FUV}}$ at both z$\sim$0 and z$\sim 0.7$ and is not highly dependent on the choice of input parameters $\tau_B^f$ and $F$. 

We find that the SFR correction methods tested give similar results, in that they perform similarly well, for both the full sample of star-forming galaxies and our restricted samples (with a cut for larger sizes and lower S\'{e}rsic indexes) of large disc-dominated galaxies. 
One difference between the samples is that the dispersion is smaller for the restricted sample at z$\sim$0. This might be because the restricted sample spans a smaller range of galaxy morphologies, only including large exponential discs. 

Despite the small number of galaxies detected in COSMOS at multiple UV wavelengths, we are able to learn something from the UV slope measurements at $z\sim0.7$.
The attenuation at higher redshift is so substantial that inclination dependency is probably not the dominant concern for the SFR$_\text{FUV}$ measurements because accurately correcting the average attenuation of the population gives the most improvement in terms of SFRs. At both z$\sim$0 and $z\sim0.7$, applying an attenuation correction based on the UV slope is better than no correction, but at $z\sim0.7,$ the SFRs on average still lie below the MS by $\sim$0.44 dex and $\sim$0.15 dex for the full and restricted samples, respectively (for a galaxy at $1-\cos(i)=0.5$).

L18 found that dust at $z\sim0.7$ is most closely associated with star-forming clumps rather than the disc. Therefore, we expect the UV attenuation curve to be relatively constant with disc inclination angle. 
In the local sample, we found that adopting a single UV-slope
attenuation relation was unable to remove the inclination bias. The attenuation for face-on galaxies is dominated by the clumpy dust component, whereas the attenuation for edge-on galaxies is more affected by the diffuse dust disc, and these two components have different UV-attenuation laws (e.g. \cite{Wild2011, daCunha2008,Charlot2000}). Therefore, a different A$_{FUV}$--$\beta$ relation would be required at different galaxy inclinations. 
Perhaps at the highest inclination angles, where birth-clouds might overlap along the line of sight and photons would have to escape through more optically thick material, we might see a drop-off in FUV light. A cross-over of star-forming clouds would depend on the cloud number density. 
Galaxies around cosmic noon ($z\sim2$) were forming stars at a faster rate, having more birth clouds. On one hand, the higher SFR would increase the probability that the birth clouds overlap along the line of sight for inclined galaxies. On the other hand, the scale-height of the gas disc is also expected to be larger at this redshift \citep{ForsterSchreiber2009,Kassin2012,Stott2016,Turner2017}. A larger scale-height would dampen any expected inclination dependency. Therefore it would be interesting to repeat this exercise at $z=1.5$ in the era of the James Webb Space Telescope, where high-resolution rest-frame optical imaging can provide morphology and inclination measurements of massive galaxies.

\begin{acknowledgements}
SL acknowledges funding from Deutsche Forschungsgemeinschaft (DFG) Grant BE 1837 / 13-1 r. ES and PL acknowledge funding from the European Research Council (ERC) under the European Union’s Horizon 2020 research and innovation programme (grant agreement No. 694343). MTS acknowledges support from a Royal Society Leverhulme Trust Senior Research Fellowship (LT150041). BG acknowledges the support of the Australian Research Council as the recipient of a Future Fellowship (FT140101202). Thank you to the referee and editors for improving this manuscript. 

\end{acknowledgements}
\bibliographystyle{aa} 
  \bibliography{inclination} 

\begin{thebibliography}{61}
\expandafter\ifx\csname natexlab\endcsname\relax\def\natexlab#1{#1}\fi

\bibitem[{{Allamandola} {et~al.}(1985){Allamandola}, {Tielens}, \&
  {Barker}}]{Allamandola1985}
{Allamandola}, L.~J., {Tielens}, A.~G.~G.~M., \& {Barker}, J.~R. 1985, \apjl,
  290, L25

\bibitem[{{Battisti} {et~al.}(2016){Battisti}, {Calzetti}, \&
  {Chary}}]{Battisti2016}
{Battisti}, A.~J., {Calzetti}, D., \& {Chary}, R.-R. 2016, \apj, 818, 13

\bibitem[{{Bendo} {et~al.}(2010){Bendo}, {Wilson}, {Pohlen}, {Sauvage}, {Auld},
  {Baes}, {Barlow}, {Bock}, {Boselli}, {Bradford}, {Buat}, {Castro-Rodriguez},
  {Chanial}, {Charlot}, {Ciesla}, {Clements}, {Cooray}, {Cormier}, {Cortese},
  {Davies}, {Dwek}, {Eales}, {Elbaz}, {Galametz}, {Galliano}, {Gear}, {Glenn},
  {Gomez}, {Griffin}, {Hony}, {Isaak}, {Levenson}, {Lu}, {Madden},
  {O'Halloran}, {Okumura}, {Oliver}, {Page}, {Panuzzo}, {Papageorgiou},
  {Parkin}, {Perez-Fournon}, {Rangwala}, {Rigby}, {Roussel}, {Rykala},
  {Sacchi}, {Schulz}, {Schirm}, {Smith}, {Spinoglio}, {Stevens}, {Sundar},
  {Symeonidis}, {Trichas}, {Vaccari}, {Vigroux}, {Wozniak}, {Wright}, \&
  {Zeilinger}}]{Bendo2010}
{Bendo}, G.~J., {Wilson}, C.~D., {Pohlen}, M., {et~al.} 2010, \aap, 518, L65

\bibitem[{{Bianchi} {et~al.}(2011){Bianchi}, {Efremova}, {Herald}, {Girardi},
  {Zabot}, {Marigo}, \& {Martin}}]{Bianchi2011}
{Bianchi}, L., {Efremova}, B., {Herald}, J., {et~al.} 2011, \mnras, 411, 2770

\bibitem[{{Boquien} {et~al.}(2012){Boquien}, {Buat}, {Boselli}, {Baes},
  {Bendo}, {Ciesla}, {Cooray}, {Cortese}, {Eales}, {Gavazzi}, {Gomez},
  {Lebouteiller}, {Pappalardo}, {Pohlen}, {Smith}, \&
  {Spinoglio}}]{Boquien2012}
{Boquien}, M., {Buat}, V., {Boselli}, A., {et~al.} 2012, \aap, 539, A145

\bibitem[{{Boquien} {et~al.}(2016){Boquien}, {Kennicutt}, {Calzetti}, {Dale},
  {Galametz}, {Sauvage}, {Croxall}, {Draine}, {Kirkpatrick}, {Kumari}, {Hunt},
  {De Looze}, {Pellegrini}, {Rela{\~n}o}, {Smith}, \&
  {Tabatabaei}}]{Boquien2016}
{Boquien}, M., {Kennicutt}, R., {Calzetti}, D., {et~al.} 2016, \aap, 591, A6

\bibitem[{{Buat} {et~al.}(1999){Buat}, {Donas}, {Milliard}, \& {Xu}}]{Buat1999}
{Buat}, V., {Donas}, J., {Milliard}, B., \& {Xu}, C. 1999, \aap, 352, 371

\bibitem[{{Calzetti}(1997)}]{Calzetti1997}
{Calzetti}, D. 1997, \aj, 113, 162

\bibitem[{{Calzetti}(2013)}]{Calzetti2013}
{Calzetti}, D. 2013, {Star Formation Rate Indicators}, ed.
  J.~{Falc{\'o}n-Barroso} \& J.~H. {Knapen}, 419

\bibitem[{{Calzetti} {et~al.}(1994){Calzetti}, {Kinney}, \&
  {Storchi-Bergmann}}]{Calzetti1994}
{Calzetti}, D., {Kinney}, A.~L., \& {Storchi-Bergmann}, T. 1994, \apj, 429, 582

\bibitem[{{Cardelli} {et~al.}(1989){Cardelli}, {Clayton}, \&
  {Mathis}}]{Cardelli1989}
{Cardelli}, J.~A., {Clayton}, G.~C., \& {Mathis}, J.~S. 1989, \apj, 345, 245

\bibitem[{{Catal{\'a}n-Torrecilla} {et~al.}(2015){Catal{\'a}n-Torrecilla}, {Gil
  de Paz}, {Castillo-Morales}, {Iglesias-P{\'a}ramo}, {S{\'a}nchez},
  {Kennicutt}, {P{\'e}rez-Gonz{\'a}lez}, {Marino}, {Walcher}, {Husemann},
  {Garc{\'{\i}}a-Benito}, {Mast}, {Gonz{\'a}lez Delgado}, {Mu{\~n}oz-Mateos},
  {Bland-Hawthorn}, {Bomans}, {Del Olmo}, {Galbany}, {Gomes}, {Kehrig},
  {L{\'o}pez-S{\'a}nchez}, {Mendoza}, {Monreal-Ibero}, {P{\'e}rez-Torres},
  {S{\'a}nchez-Bl{\'a}zquez}, {Vilchez}, \& {Califa
  Collaboration}}]{CatalanTorrecilla2015}
{Catal{\'a}n-Torrecilla}, C., {Gil de Paz}, A., {Castillo-Morales}, A.,
  {et~al.} 2015, \aap, 584, A87

\bibitem[{{Chabrier}(2003)}]{Chabrier2003}
{Chabrier}, G. 2003, \pasp, 115, 763

\bibitem[{{Charlot} \& {Fall}(2000)}]{Charlot2000}
{Charlot}, S. \& {Fall}, S.~M. 2000, \apj, 539, 718

\bibitem[{{Cluver} {et~al.}(2017){Cluver}, {Jarrett}, {Dale}, {Smith},
  {August}, \& {Brown}}]{Cluver2017}
{Cluver}, M.~E., {Jarrett}, T.~H., {Dale}, D.~A., {et~al.} 2017, \apj, 850, 68

\bibitem[{{Cluver} {et~al.}(2014){Cluver}, {Jarrett}, {Hopkins}, {Driver},
  {Liske}, {Gunawardhana}, {Taylor}, {Robotham}, {Alpaslan}, {Baldry}, {Brown},
  {Peacock}, {Popescu}, {Tuffs}, {Bauer}, {Bland-Hawthorn}, {Colless},
  {Holwerda}, {Lara-L{\'o}pez}, {Leschinski}, {L{\'o}pez-S{\'a}nchez},
  {Norberg}, {Owers}, {Wang}, \& {Wilkins}}]{Cluver2014}
{Cluver}, M.~E., {Jarrett}, T.~H., {Hopkins}, A.~M., {et~al.} 2014, \apj, 782,
  90

\bibitem[{{da Cunha} {et~al.}(2008){da Cunha}, {Charlot}, \&
  {Elbaz}}]{daCunha2008}
{da Cunha}, E., {Charlot}, S., \& {Elbaz}, D. 2008, \mnras, 388, 1595

\bibitem[{{Dale} {et~al.}(2017){Dale}, {Cook}, {Roussel}, {Turner}, {Armus},
  {Bolatto}, {Boquien}, {Brown}, {Calzetti}, {De Looze}, {Galametz}, {Gordon},
  {Groves}, {Jarrett}, {Helou}, {Herrera-Camus}, {Hinz}, {Hunt}, {Kennicutt},
  {Murphy}, {Rest}, {Sandstrom}, {Smith}, {Tabatabaei}, \& {Wilson}}]{Dale2017}
{Dale}, D.~A., {Cook}, D.~O., {Roussel}, H., {et~al.} 2017, \apj, 837, 90

\bibitem[{{Davies} {et~al.}(2016){Davies}, {Driver}, {Robotham}, {Grootes},
  {Popescu}, {Tuffs}, {Hopkins}, {Alpaslan}, {Andrews}, {Bland-Hawthorn},
  {Bremer}, {Brough}, {Brown}, {Cluver}, {Croom}, {da Cunha}, {Dunne},
  {Lara-L{\'o}pez}, {Liske}, {Loveday}, {Moffett}, {Owers}, {Phillipps},
  {Sansom}, {Taylor}, {Michalowski}, {Ibar}, {Smith}, \& {Bourne}}]{Davies2016}
{Davies}, L.~J.~M., {Driver}, S.~P., {Robotham}, A.~S.~G., {et~al.} 2016,
  \mnras, 461, 458

\bibitem[{{Dopita} {et~al.}(2003){Dopita}, {Groves}, {Sutherland}, \&
  {Kewley}}]{Dopita2003}
{Dopita}, M.~A., {Groves}, B.~A., {Sutherland}, R.~S., \& {Kewley}, L.~J. 2003,
  \apj, 583, 727

\bibitem[{Foreman-Mackey(2017)}]{DFM17}
Foreman-Mackey, D. 2017, Fitting a plane to data,
  \url{http://dfm.io/posts/fitting-a-plane/}

\bibitem[{{Foreman-Mackey} {et~al.}(2013){Foreman-Mackey}, {Hogg}, {Lang}, \&
  {Goodman}}]{ForemanMackey2013}
{Foreman-Mackey}, D., {Hogg}, D.~W., {Lang}, D., \& {Goodman}, J. 2013, \pasp,
  125, 306

\bibitem[{{F{\"o}rster Schreiber} {et~al.}(2009){F{\"o}rster Schreiber},
  {Genzel}, {Bouch{\'e}}, {Cresci}, {Davies}, {Buschkamp}, {Shapiro},
  {Tacconi}, {Hicks}, {Genel}, {Shapley}, {Erb}, {Steidel}, {Lutz},
  {Eisenhauer}, {Gillessen}, {Sternberg}, {Renzini}, {Cimatti}, {Daddi},
  {Kurk}, {Lilly}, {Kong}, {Lehnert}, {Nesvadba}, {Verma}, {McCracken},
  {Arimoto}, {Mignoli}, \& {Onodera}}]{ForsterSchreiber2009}
{F{\"o}rster Schreiber}, N.~M., {Genzel}, R., {Bouch{\'e}}, N., {et~al.} 2009,
  \apj, 706, 1364

\bibitem[{{Grootes} {et~al.}(2017){Grootes}, {Tuffs}, {Popescu}, {Norberg},
  {Robotham}, {Liske}, {Andrae}, {Baldry}, {Gunawardhana}, {Kelvin}, {Madore},
  {Seibert}, {Taylor}, {Alpaslan}, {Brown}, {Cluver}, {Driver},
  {Bland-Hawthorn}, {Holwerda}, {Hopkins}, {Lopez-Sanchez}, {Loveday}, \&
  {Rushton}}]{Grootes2017}
{Grootes}, M.~W., {Tuffs}, R.~J., {Popescu}, C.~C., {et~al.} 2017, \aj, 153,
  111

\bibitem[{{Grootes} {et~al.}(2013){Grootes}, {Tuffs}, {Popescu}, {Pastrav},
  {Andrae}, {Gunawardhana}, {Kelvin}, {Liske}, {Seibert}, {Taylor}, {Graham},
  {Baes}, {Baldry}, {Bourne}, {Brough}, {Cooray}, {Dariush}, {De Zotti},
  {Driver}, {Dunne}, {Gomez}, {Hopkins}, {Hopwood}, {Jarvis}, {Loveday},
  {Maddox}, {Madore}, {Micha{\l}owski}, {Norberg}, {Parkinson}, {Prescott},
  {Robotham}, {Smith}, {Thomas}, \& {Valiante}}]{Grootes2013}
{Grootes}, M.~W., {Tuffs}, R.~J., {Popescu}, C.~C., {et~al.} 2013, \apj, 766,
  59

\bibitem[{{Groves} {et~al.}(2012){Groves}, {Krause}, {Sandstrom}, {Schmiedeke},
  {Leroy}, {Linz}, {Kapala}, {Rix}, {Schinnerer}, {Tabatabaei}, {Walter}, \&
  {da Cunha}}]{Groves2012}
{Groves}, B., {Krause}, O., {Sandstrom}, K., {et~al.} 2012, \mnras, 426, 892

\bibitem[{{Hao} {et~al.}(2011){Hao}, {Kennicutt}, {Johnson}, {Calzetti},
  {Dale}, \& {Moustakas}}]{Hao2011}
{Hao}, C.-N., {Kennicutt}, R.~C., {Johnson}, B.~D., {et~al.} 2011, \apj, 741,
  124

\bibitem[{{Izotova} \& {Izotov}(2018)}]{Izotova2018}
{Izotova}, I.~Y. \& {Izotov}, Y.~I. 2018, \apss, 363, 47

\bibitem[{{Jarrett} {et~al.}(2011){Jarrett}, {Cohen}, {Masci}, {Wright},
  {Stern}, {Benford}, {Blain}, {Carey}, {Cutri}, {Eisenhardt}, {Lonsdale},
  {Mainzer}, {Marsh}, {Padgett}, {Petty}, {Ressler}, {Skrutskie}, {Stanford},
  {Surace}, {Tsai}, {Wheelock}, \& {Yan}}]{Jarrett2011}
{Jarrett}, T.~H., {Cohen}, M., {Masci}, F., {et~al.} 2011, \apj, 735, 112

\bibitem[{{Kassin} {et~al.}(2012){Kassin}, {Weiner}, {Faber}, {Gardner},
  {Willmer}, {Coil}, {Cooper}, {Devriendt}, {Dutton}, {Guhathakurta}, {Koo},
  {Metevier}, {Noeske}, \& {Primack}}]{Kassin2012}
{Kassin}, S.~A., {Weiner}, B.~J., {Faber}, S.~M., {et~al.} 2012, \apj, 758, 106

\bibitem[{{Kennicutt} \& {Evans}(2012)}]{Kennicutt2012}
{Kennicutt}, R.~C. \& {Evans}, N.~J. 2012, \araa, 50, 531

\bibitem[{{Kennicutt} {et~al.}(2009){Kennicutt}, {Hao}, {Calzetti},
  {Moustakas}, {Dale}, {Bendo}, {Engelbracht}, {Johnson}, \&
  {Lee}}]{Kennicutt2009}
{Kennicutt}, Jr., R.~C., {Hao}, C.-N., {Calzetti}, D., {et~al.} 2009, \apj,
  703, 1672

\bibitem[{{Kroupa}(2001)}]{Kroupa2001}
{Kroupa}, P. 2001, \mnras, 322, 231

\bibitem[{{Laigle} {et~al.}(2016){Laigle}, {McCracken}, {Ilbert}, {Hsieh},
  {Davidzon}, {Capak}, {Hasinger}, {Silverman}, {Pichon}, {Coupon}, {Aussel},
  {Le Borgne}, {Caputi}, {Cassata}, {Chang}, {Civano}, {Dunlop}, {Fynbo},
  {Kartaltepe}, {Koekemoer}, {Le F{\`e}vre}, {Le Floc'h}, {Leauthaud}, {Lilly},
  {Lin}, {Marchesi}, {Milvang-Jensen}, {Salvato}, {Sanders}, {Scoville},
  {Smolcic}, {Stockmann}, {Taniguchi}, {Tasca}, {Toft}, {Vaccari}, \&
  {Zabl}}]{Laigle2016}
{Laigle}, C., {McCracken}, H.~J., {Ilbert}, O., {et~al.} 2016, \apjs, 224, 24

\bibitem[{{Larson} {et~al.}(2011){Larson}, {Dunkley}, {Hinshaw}, {Komatsu},
  {Nolta}, {Bennett}, {Gold}, {Halpern}, {Hill}, {Jarosik}, {Kogut}, {Limon},
  {Meyer}, {Odegard}, {Page}, {Smith}, {Spergel}, {Tucker}, {Weiland},
  {Wollack}, \& {Wright}}]{Larson2011}
{Larson}, D., {Dunkley}, J., {Hinshaw}, G., {et~al.} 2011, \apjs, 192, 16

\bibitem[{{Leger} \& {Puget}(1984)}]{Leger1984}
{Leger}, A. \& {Puget}, J.~L. 1984, \aap, 137, L5

\bibitem[{{Leitherer} {et~al.}(1999){Leitherer}, {Schaerer}, {Goldader},
  {Delgado}, {Robert}, {Kune}, {de Mello}, {Devost}, \&
  {Heckman}}]{Leitherer1999}
{Leitherer}, C., {Schaerer}, D., {Goldader}, J.~D., {et~al.} 1999, \apjs, 123,
  3

\bibitem[{{Leitherer} {et~al.}(2011){Leitherer}, {Tremonti}, {Heckman}, \&
  {Calzetti}}]{Leitherer2011}
{Leitherer}, C., {Tremonti}, C.~A., {Heckman}, T.~M., \& {Calzetti}, D. 2011,
  \aj, 141, 37

\bibitem[{{Leslie} {et~al.}(2018){Leslie}, {Sargent}, {Schinnerer}, {Groves},
  {van der Wel}, {Zamorani}, {Fudamoto}, {Lang}, \& {Smol{\v
  c}i{\'c}}}]{Leslie2017}
{Leslie}, S.~K., {Sargent}, M.~T., {Schinnerer}, E., {et~al.} 2018, ArXiv
  e-prints [\eprint[arXiv]{1801.03501}]

\bibitem[{{Madau} \& {Dickinson}(2014)}]{Madau2014}
{Madau}, P. \& {Dickinson}, M. 2014, \araa, 52, 415

\bibitem[{{Magnelli} {et~al.}(2014){Magnelli}, {Lutz}, {Saintonge}, {Berta},
  {Santini}, {Symeonidis}, {Altieri}, {Andreani}, {Aussel}, {B{\'e}thermin},
  {Bock}, {Bongiovanni}, {Cepa}, {Cimatti}, {Conley}, {Daddi}, {Elbaz},
  {F{\"o}rster Schreiber}, {Genzel}, {Ivison}, {Le Floc'h}, {Magdis},
  {Maiolino}, {Nordon}, {Oliver}, {Page}, {P{\'e}rez Garc{\'{\i}}a},
  {Poglitsch}, {Popesso}, {Pozzi}, {Riguccini}, {Rodighiero}, {Rosario},
  {Roseboom}, {Sanchez-Portal}, {Scott}, {Sturm}, {Tacconi}, {Valtchanov},
  {Wang}, \& {Wuyts}}]{Magnelli2014}
{Magnelli}, B., {Lutz}, D., {Saintonge}, A., {et~al.} 2014, \aap, 561, A86

\bibitem[{{Martin} {et~al.}(2005){Martin}, {Fanson}, {Schiminovich},
  {Morrissey}, {Friedman}, {Barlow}, {Conrow}, {Grange}, {Jelinsky},
  {Milliard}, {Siegmund}, {Bianchi}, {Byun}, {Donas}, {Forster}, {Heckman},
  {Lee}, {Madore}, {Malina}, {Neff}, {Rich}, {Small}, {Surber}, {Szalay},
  {Welsh}, \& {Wyder}}]{Martin2005}
{Martin}, D.~C., {Fanson}, J., {Schiminovich}, D., {et~al.} 2005, \apjl, 619,
  L1

\bibitem[{{Meurer} {et~al.}(1999){Meurer}, {Heckman}, \&
  {Calzetti}}]{Meurer1999}
{Meurer}, G.~R., {Heckman}, T.~M., \& {Calzetti}, D. 1999, \apj, 521, 64

\bibitem[{{Murphy} {et~al.}(2011){Murphy}, {Condon}, {Schinnerer}, {Kennicutt},
  {Calzetti}, {Armus}, {Helou}, {Turner}, {Aniano}, {Beir{\~a}o}, {Bolatto},
  {Brandl}, {Croxall}, {Dale}, {Donovan Meyer}, {Draine}, {Engelbracht},
  {Hunt}, {Hao}, {Koda}, {Roussel}, {Skibba}, \& {Smith}}]{Murphy2011}
{Murphy}, E.~J., {Condon}, J.~J., {Schinnerer}, E., {et~al.} 2011, \apj, 737,
  67

\bibitem[{{Natta} \& {Panagia}(1984)}]{Natta1984}
{Natta}, A. \& {Panagia}, N. 1984, \apj, 287, 228

\bibitem[{{Overzier} {et~al.}(2011){Overzier}, {Heckman}, {Wang}, {Armus},
  {Buat}, {Howell}, {Meurer}, {Seibert}, {Siana}, {Basu-Zych}, {Charlot},
  {Gon{\c c}alves}, {Martin}, {Neill}, {Rich}, {Salim}, \&
  {Schiminovich}}]{Overzier2011}
{Overzier}, R.~A., {Heckman}, T.~M., {Wang}, J., {et~al.} 2011, \apjl, 726, L7

\bibitem[{{Popescu} {et~al.}(2011){Popescu}, {Tuffs}, {Dopita}, {Fischera},
  {Kylafis}, \& {Madore}}]{Popescu2011}
{Popescu}, C.~C., {Tuffs}, R.~J., {Dopita}, M.~A., {et~al.} 2011, \aap, 527,
  A109

\bibitem[{{Salmon} {et~al.}(2016){Salmon}, {Papovich}, {Long}, {Willner},
  {Finkelstein}, {Ferguson}, {Dickinson}, {Duncan}, {Faber}, {Hathi},
  {Koekemoer}, {Kurczynski}, {Newman}, {Pacifici}, {P{\'e}rez-Gonz{\'a}lez}, \&
  {Pforr}}]{Salmon2016}
{Salmon}, B., {Papovich}, C., {Long}, J., {et~al.} 2016, \apj, 827, 20

\bibitem[{{Santini} {et~al.}(2014){Santini}, {Maiolino}, {Magnelli}, {Lutz},
  {Lamastra}, {Li Causi}, {Eales}, {Andreani}, {Berta}, {Buat}, {Cooray},
  {Cresci}, {Daddi}, {Farrah}, {Fontana}, {Franceschini}, {Genzel}, {Granato},
  {Grazian}, {Le Floc'h}, {Magdis}, {Magliocchetti}, {Mannucci}, {Menci},
  {Nordon}, {Oliver}, {Popesso}, {Pozzi}, {Riguccini}, {Rodighiero}, {Rosario},
  {Salvato}, {Scott}, {Silva}, {Tacconi}, {Viero}, {Wang}, {Wuyts}, \&
  {Xu}}]{Santini2014}
{Santini}, P., {Maiolino}, R., {Magnelli}, B., {et~al.} 2014, \aap, 562, A30

\bibitem[{{Sargent} {et~al.}(2007){Sargent}, {Carollo}, {Lilly}, {Scarlata},
  {Feldmann}, {Kampczyk}, {Koekemoer}, {Scoville}, {Kneib}, {Leauthaud},
  {Massey}, {Rhodes}, {Tasca}, {Capak}, {McCracken}, {Porciani}, {Renzini},
  {Taniguchi}, {Thompson}, \& {Sheth}}]{Sargent2007}
{Sargent}, M.~T., {Carollo}, C.~M., {Lilly}, S.~J., {et~al.} 2007, \apjs, 172,
  434

\bibitem[{{Sargent} {et~al.}(2014){Sargent}, {Daddi}, {B{\'e}thermin},
  {Aussel}, {Magdis}, {Hwang}, {Juneau}, {Elbaz}, \& {da Cunha}}]{Sargent2014}
{Sargent}, M.~T., {Daddi}, E., {B{\'e}thermin}, M., {et~al.} 2014, \apj, 793,
  19

\bibitem[{{Simard} {et~al.}(2011){Simard}, {Mendel}, {Patton}, {Ellison}, \&
  {McConnachie}}]{Simard2011}
{Simard}, L., {Mendel}, J.~T., {Patton}, D.~R., {Ellison}, S.~L., \&
  {McConnachie}, A.~W. 2011, \apjs, 196, 11

\bibitem[{{Smith} {et~al.}(2007){Smith}, {Draine}, {Dale}, {Moustakas},
  {Kennicutt}, {Helou}, {Armus}, {Roussel}, {Sheth}, {Bendo}, {Buckalew},
  {Calzetti}, {Engelbracht}, {Gordon}, {Hollenbach}, {Li}, {Malhotra},
  {Murphy}, \& {Walter}}]{Smith2007}
{Smith}, J.~D.~T., {Draine}, B.~T., {Dale}, D.~A., {et~al.} 2007, \apj, 656,
  770

\bibitem[{{Stott} {et~al.}(2016){Stott}, {Swinbank}, {Johnson}, {Tiley},
  {Magdis}, {Bower}, {Bunker}, {Bureau}, {Harrison}, {Jarvis}, {Sharples},
  {Smail}, {Sobral}, {Best}, \& {Cirasuolo}}]{Stott2016}
{Stott}, J.~P., {Swinbank}, A.~M., {Johnson}, H.~L., {et~al.} 2016, \mnras,
  457, 1888

\bibitem[{{Treyer} {et~al.}(2010){Treyer}, {Schiminovich}, {Johnson}, {O'Dowd},
  {Martin}, {Wyder}, {Charlot}, {Heckman}, {Martins}, {Seibert}, \& {van der
  Hulst}}]{Treyer2010}
{Treyer}, M., {Schiminovich}, D., {Johnson}, B.~D., {et~al.} 2010, \apj, 719,
  1191

\bibitem[{{Tuffs} {et~al.}(2004){Tuffs}, {Popescu}, {V{\"o}lk}, {Kylafis}, \&
  {Dopita}}]{Tuffs2004}
{Tuffs}, R.~J., {Popescu}, C.~C., {V{\"o}lk}, H.~J., {Kylafis}, N.~D., \&
  {Dopita}, M.~A. 2004, \aap, 419, 821

\bibitem[{{Turner} {et~al.}(2017){Turner}, {Cirasuolo}, {Harrison}, {McLure},
  {Dunlop}, {Swinbank}, {Johnson}, {Sobral}, {Matthee}, \&
  {Sharples}}]{Turner2017}
{Turner}, O.~J., {Cirasuolo}, M., {Harrison}, C.~M., {et~al.} 2017, \mnras,
  471, 1280

\bibitem[{{Wang} {et~al.}(2016){Wang}, {Norberg}, {Gunawardhana}, {Heinis},
  {Baldry}, {Bland-Hawthorn}, {Bourne}, {Brough}, {Brown}, {Cluver}, {Cooray},
  {da Cunha}, {Driver}, {Dunne}, {Dye}, {Eales}, {Grootes}, {Holwerda},
  {Hopkins}, {Ibar}, {Ivison}, {Lacey}, {Lara-Lopez}, {Loveday}, {Maddox},
  {Micha{\l}owski}, {Oteo}, {Owers}, {Popescu}, {Smith}, {Taylor}, {Tuffs}, \&
  {van der Werf}}]{Wang2016}
{Wang}, L., {Norberg}, P., {Gunawardhana}, M.~L.~P., {et~al.} 2016, \mnras,
  461, 1898

\bibitem[{{Wild} {et~al.}(2011){Wild}, {Charlot}, {Brinchmann}, {Heckman},
  {Vince}, {Pacifici}, \& {Chevallard}}]{Wild2011}
{Wild}, V., {Charlot}, S., {Brinchmann}, J., {et~al.} 2011, \mnras, 417, 1760

\bibitem[{{Wuyts} {et~al.}(2011){Wuyts}, {F{\"o}rster Schreiber}, {Lutz},
  {Nordon}, {Berta}, {Altieri}, {Andreani}, {Aussel}, {Bongiovanni}, {Cepa},
  {Cimatti}, {Daddi}, {Elbaz}, {Genzel}, {Koekemoer}, {Magnelli}, {Maiolino},
  {McGrath}, {P{\'e}rez Garc{\'{\i}}a}, {Poglitsch}, {Popesso}, {Pozzi},
  {Sanchez-Portal}, {Sturm}, {Tacconi}, \& {Valtchanov}}]{Wuyts2011a}
{Wuyts}, S., {F{\"o}rster Schreiber}, N.~M., {Lutz}, D., {et~al.} 2011, \apj,
  738, 106

\bibitem[{{Wuyts} {et~al.}(2008){Wuyts}, {Labb{\'e}}, {F{\"o}rster Schreiber},
  {Franx}, {Rudnick}, {Brammer}, \& {van Dokkum}}]{Wuyts2008}
{Wuyts}, S., {Labb{\'e}}, I., {F{\"o}rster Schreiber}, N.~M., {et~al.} 2008,
  \apj, 682, 985

\end{thebibliography}
%

%
\appendix

\section{Different UV-slope-related corrections at $z\sim0.7$ }\label{A1}

\begin{figure}
\includegraphics[width = \linewidth ]{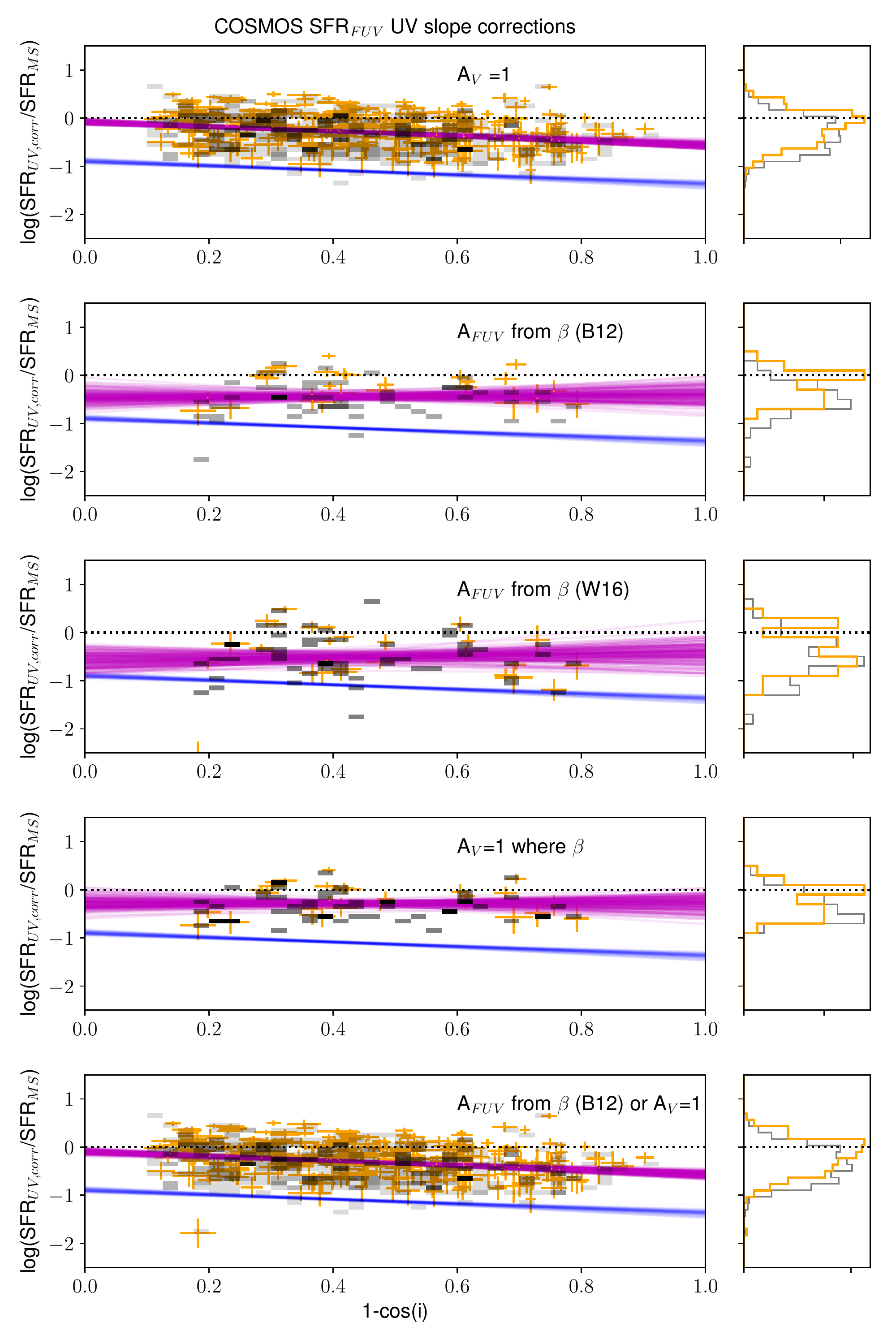}
\caption{SFR$_{FUV,corr}$ vs. inclination for different UV-slope-related correction methods, as described in Appendix \ref{A1}. }\label{figA1}
\end{figure}

In Section \ref{uvslope}, our ability to test the performance of using the galaxy rest-frame UV slope $\beta$ to correct attenuated FUV SFRs was limited for the COSMOS $\sim0.7$ sample by the small number of galaxies (56 in the full and 28 in the restricted samples). In this section, we test the performance of the assumption that A$_\text{FUV} = 2.035$ (which can be applied to all galaxies) and the use of a different $\beta$--A$_\text{FUV}$ relation. The results of these tests are summarised in Table \ref{tabA1} and Figure \ref{figA1}. We find no obvious difference in the performance of SFR$_\text{FUV}$ attenuation corrections assuming the \cite{Wang2016} or \cite{Boquien2012} $\beta$--A$_\text{FUV}$ relation or a constant $A_V=1$ for our limited sample of galaxies. 

To build a statistically significant sample of galaxies in COSMOS at $z\sim0.7,$ we followed a standard assumption and assumed $A_V=1$. First, to judge whether assuming $A_V=1$ performs better than $\beta$ at correcting the systematic SFR$_\text{FUV}$/SFR$_\text{MS}$ offset, we have calculated SFR$_\text{FUV,corr}$ assuming $A_V=1$ for the 56 (28) galaxies with determined UV-slopes in the full (restricted) samples. We assumed the average Milky Way attenuation curve of \cite{Cardelli1989} and an $R_V=3.1$, resulting in an attenuation of 1 magnitude in the V band corresponding to a FUV attenuation of A$_\text{FUV} = 2.035$ (or a y-intercept offset of $\sim$1.2 dex). After correcting the galaxies with fitted $\beta$ values by A$_\text{FUV} = 2.035$, the best fits for the log(SFR$_\text{FUV,corr}$/SFR$_\text{MS}$)--inclination relation are ($k_1,k_2,\sigma^2$) = (-0.03$^{+0.40}_{-0.39}$, 0.27 $^{+0.18}_{-0.18}$,
and 0.24$^{+0.02}_{-0.03}$) for the full sample and ($k_1,k_2,\sigma^2$) = (-0.05$^{+0.50}_{-0.51}$, -0.12$^{+0.25}_{-0.24}$, and 0.25$^{+0.07}_{-0.09}$) for the restricted sample. 
Interestingly, the slope has improved (SFRs have become more inclination independent) compared to no correction, and the y-intercept is also closer to zero. Compared to the $\beta$ corrections from \cite{Boquien2012} used in Section \ref{uvslope}, the slopes are consistent given the large uncertainties, and the y-intercept has not improved at all for the restricted sample, with SFRs remaining 0.12$\pm0.25$dex below the MS. However, the small sample size of galaxies with $\beta$ still inhibits a robust comparison.

Applying the correction $A_V=1$ to all the galaxies in the COSMOS sample results in best-fit values of ($k_1,k_2,\sigma^2$) = (-0.48$^{+0.12}_{-0.12}$, -0.08$^{+0.06}_{-0.05}$, and 0.28$^{+0.02}_{-0.03}$) for the full sample and ($k_1,k_2,\sigma^2$) = (-0.44$^{+0.15}_{-0.15}$, 0.06$^{+0.07}_{-0.07}$, and 0.26$^{+0.02}_{-0.02}$) for the restricted sample.
As expected, the slope and scatter remain the same as for the uncorrected case, and the y-intercept becomes more consistent with zero. This indicates that using $A_\text{FUV}=2.035$ is an appropriate method for correcting the average dust attenuation for our sample of massive galaxies at $z\sim0.7$, but a larger $A_\text{FUV}$ is required to bring the average SFR in line with the expected MS SFR for the full sample; a galaxy in the full sample or restricted sample with 1-cos($i$)=0.5 lies 0.32 or 0.16 dex below the MS, respectively. 

Combining the two approaches, we also fitted our samples adopting the UV-slope-derived attenuation where available and adopting A$_\text{FUV} = 2.035$ where $\beta$ was unavailable. 
Applying attenuation corrections based on the UV-slope or $A_V=1$ at $z\sim0.7$ does not remove the inclination dependency of the SFR$_\text{FUV}$: $k_1=-0.48\pm0.12$ and $-0.47\pm0.16$ for the full and restricted samples, respectively. These results would suggest that applying $A_V=1$ performs similarly to the $\beta$--A$_\text{FUV}$ relation from \cite{Boquien2012} at $z\sim0.7$, but larger samples of galaxies with well-determined UV-slopes would be required to confirm this. In addition, further corrections would still be required to remove the inclination dependence of the SFRs when applying a constant A$_\text{FUV}$ correction. 

We found that Equation \ref{beta} from \cite{Boquien2016} may not be appropriate for our sample of massive disc galaxies at $z\sim0.7$ because the resulting SFRs lie below the expected MS, therefore we also tested the relation from \cite{Wang2016}:
\begin{equation}
A_\text{FUV}= 1.58\quad \beta+2.62.
\end{equation}
Comparing rows 2 and 3 of Figure \ref{figA1}, it appears that the \cite{Wang2016} and \cite{Boquien2012} corrections perform similarly. The key difference is the increased scatter in the \cite{Wang2016} SFRs due to the stronger dependence on $\beta$. Further progress in determining the attenuation law for these galaxies at intermediate redshift could be made by 1) improving our measurements of $\beta$ by using rest-frame UV spectra or more photometric observations, and 2) by increasing our sample size.

\begin{table}
\caption{COSMOS $z\sim0.7$ UV slope corrections or $A_V=1$. Fits for relations indicated in Figure \ref{figA1}. The top half shows
the full sample and the bottom half the restricted sample. The ``No corr'' and and $\beta$ (B12) methods are repeats of rows 1 and 2 in Table 1 to facilitate comparison.}\label{tabA1}
\begin{tabular}{lcccc}
\hline
Method & N & $k_1$ & $k_2$ & $\sigma^2$ \\
\hline 
No corr & 575 & -0.47 $^{+0.12 }_{-0.12 }$ & -0.89 $^{+0.06 }_{-0.05 }$ & 0.28 $^{+0.02 }_{-0.02 }$ \\
$A_V=1$&575 & -0.48 $^{+0.12 }_{-0.12 }$ & -0.08 $^{+0.06 }_{-0.05 }$ & 0.28 $^{+0.02 }_{-0.02 }$ \\
$\beta$ (B12)& 56& 0.05 $^{+0.45 }_{-0.45 }$ & -0.46 $^{+0.21 }_{-0.20 }$ & 0.29 $^{+0.06 }_{-0.07 }$ \\
$\beta$ (W16)& 56& 0.12$^{+0.62 }_{-0.60 }$ & -0.56 $^{+0.29 }_{-0.28 }$ & 0.48 $^{+0.08 }_{-0.10 }$ \\
A$_V$=1 where $\beta$&56& -0.03 $^{+0.40 }_{-0.39 }$ & -0.27 $^{+0.18 }_{-0.18 }$ & 0.24 $^{+0.05 }_{-0.06 }$ \\
$\beta$ (B12) or A$_V$=1 & 575 & -0.48 $^{+0.12 }_{-0.12 }$ & -0.10 $^{+0.06 }_{-0.06 }$ & 0.29 $^{+0.02 }_{-0.02 }$ \\
\hline
No corr& 295 &-0.44 $^{+0.15 }_{-0.15 }$ & -0.75 $^{+0.07 }_{-0.07 }$ & 0.26 $^{+0.02 }_{-0.02 }$ \\
A$_V$=1& 295& -0.44 $^{+0.15 }_{-0.15 }$ & 0.06 $^{+0.07 }_{-0.07 }$ & 0.26 $^{+0.02 }_{-0.02 }$ \\
$\beta$ (B12)&28&  -0.05 $^{+0.51 }_{-0.50 }$ & -0.12 $^{+0.25 }_{-0.24 }$ & 0.25 $^{+0.07 }_{-0.09 }$ \\
$\beta$ (W16)& 28& -0.16 $^{+0.77 }_{-0.81 }$ & -0.36 $^{+0.40 }_{-0.41 }$ & 0.53 $^{+0.12 }_{-0.17 }$ \\
A$_V$=1 where $\beta$&28&-0.05 $^{+0.50 }_{-0.51 }$ & -0.12 $^{+0.25 }_{-0.24 }$ & 0.25 $^{+0.07 }_{-0.09 }$ \\
$\beta$ (B12) or A$_V$=1 &295& -0.47 $^{+0.16 }_{-0.16 }$ & 0.06 $^{+0.07 }_{-0.07 }$ & 0.27 $^{+0.02 }_{-0.03 }$ \\
\hline
\end{tabular}
\end{table}

\section{Different hybrid UV+MIR dust corrections at $z\sim0$}\label{A2}

Using different values for $\eta$ in Equation 4 or different monochromatic L$_\text{TIR}$ conversions will naturally result in different hybrid SFR$_\text{FUV,corr}$ values. In this section, we test MIR+UV hybrid methods that use the WISE 12$\mu$m and WISE 22$\mu$m fluxes to correct for FUV attenuation in our local samples.

\begin{figure}
\includegraphics[width = \linewidth]{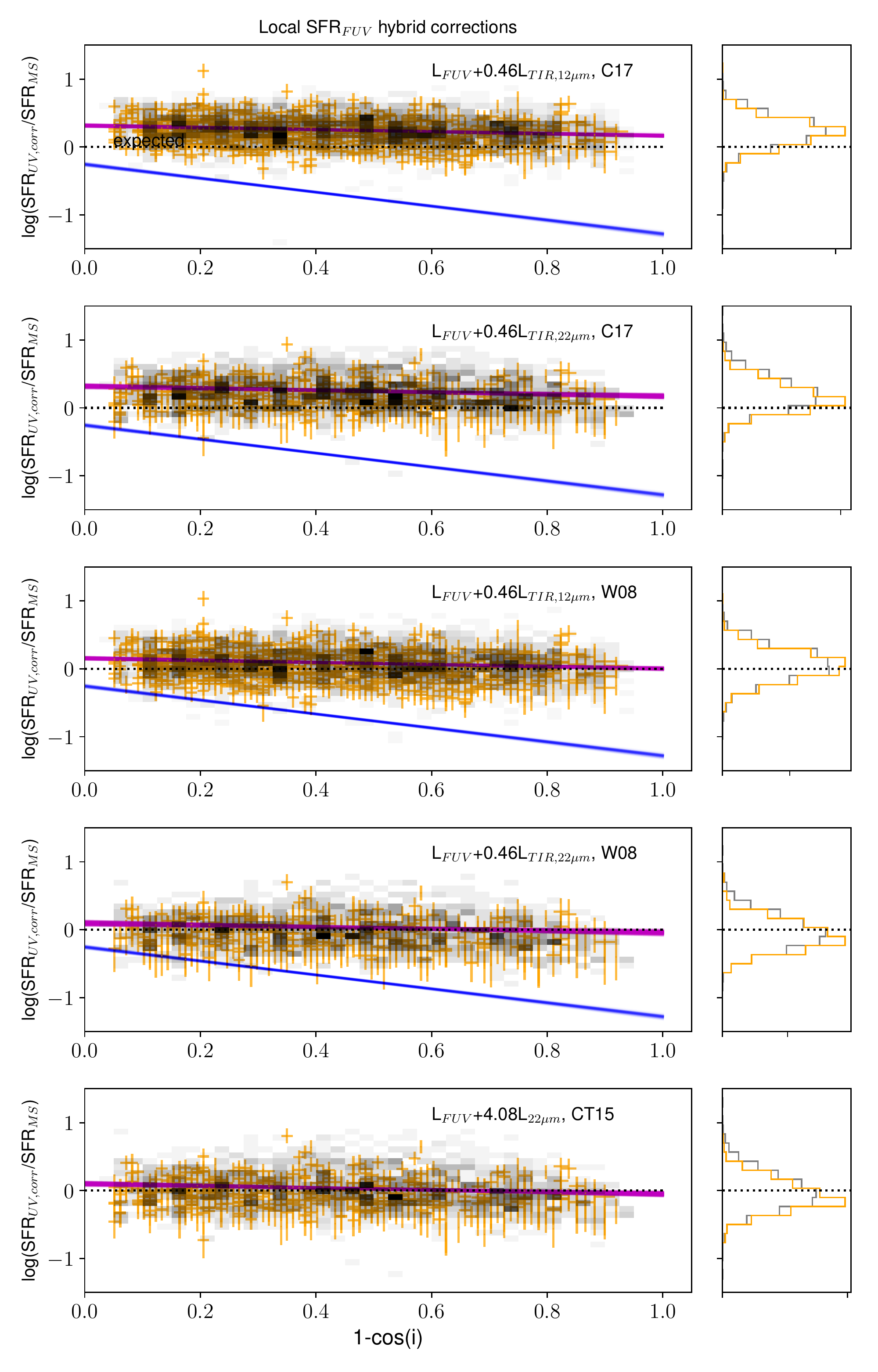}
\caption{SFR$_\text{FUV,corr}$ vs. inclination for different MIR hybrid correction methods, as indicated in the panels. C17 refers to \cite{Cluver2017}, W08 to \cite{Wuyts2008}, and CT15 to \cite{CatalanTorrecilla2015}.}\label{appendixfig}
\end{figure}

We used the relation of \citet[C17,]{Cluver2017} to convert the
W3 spectral luminosity into total-infrared luminosity (5--1100$\mu$m): \begin{equation} \log(L_{\text{IR}})=0.889\log( L_{12})+2.21,\label{l12} \end{equation} where $L_{12}$ is in units of solar luminosities. 
This relation was calibrated on the SINGS/KINGFISH sample \citep{Dale2017} supplemented with several more luminous galaxies from the literature. \cite{Cluver2017} found a 1$\sigma$ scatter of 0.15 dex in Equation \ref{l12}. 
Correcting the FUV luminosity using the \cite{Cluver2017} relation and Equation 6 results in SFR$_\text{FUV,corr}$ that are systematically $\sim$0.3 dex above the expected MS (see Table \ref{tabB1}) for both the restricted and full samples.
This may indicate that 1) a lower value of $\eta$ is required for our sample, 2) the \cite{Cluver2017} relation might give higher L$_\text{TIR}$ luminosities for our sample of galaxies because different WISE flux measurement methods were used in their study and ours, as well as a different definition of L$_{IR}$ , or 3) the adopted MS relation is $\sim$0.1 dex too high at $z\sim 0.07$.
If we had used the 22$\mu$m (W4) data rather than the 12$\mu$m (W3) fluxes to calculate L$_\text{TIR}$, and the W4 calibration from \cite{Cluver2017}, then our UV+MIR SFRs would be only slightly different, with a larger intrinsic scatter, as shown in Figure \ref{appendixfig} and Table \ref{tabB1}. \cite{Cluver2017} also found a larger scatter in the W4--TIR relation ($\sigma$ =0.18 dex)  than the W3--TIR relation ($\sigma = 0.15$). 
The uncertainty in the MS from using different SFR calibrations is about 0.1-0.2 dex, which is not sufficient to explain the 0.3 dex offset seen by the hybrid SFRs using the \cite{Cluver2017} relation for TIR luminosity. However, it is clear that the different assumptions that go in to calculating the monochromatic infrared luminosity and combining this with the FUV luminosity can alter the SFRs by $\sim$0.3 dex. 

We also tested hybrid SFRs using the \cite{Wuyts2008} SED (W08), following the methods used in L18 for both W3  (used in Section \ref{hybrid}) and W4 fluxes. These methods give SFRs closer to the expected MS values, $k_2=0.15$ and 0.14 for the 12$\mu$m (W3) full and restricted samples, respectively, and $k_2=0.1$ and 0.05 for the 22$\mu$m (W4) full and restricted samples, respectively. However, the slope of the relation is still $\sim$-0.15 dex for the full sample. The \cite{Wuyts2008} W4 correction on the restricted sample is the only hybrid example that shows a positive correlation between W4-corrected SFRs and inclination ($k_1=0.18\pm0.1$), but this is also the sub-sample with the fewest detected galaxies.  

Rather than using a flux density to LIR conversion, \citet[CT15]{CatalanTorrecilla2015}, gives the following prediction based on the energy balance approach:
\begin{equation}
L_\text{FUV,corr} = L_\text{FUV} + 4.08L_{22\mu m}.
\end{equation}
This correction results in SFRs similar to the W08 W4 hybrid SFRs. A galaxy in the full sample with inclination 1-cos(i)=0.5 lies 0.025 dex above the MS. On the other hand, a galaxy in the restricted sample at 1-cos($i$)=0.5 lies -0.05 dex below the MS. \cite{CatalanTorrecilla2015} found a coefficient of $\sim$4.5 (rather than 4.08 from the energy balance argument) when fitting to star-forming CALIFA galaxies, which would again increase the SFRs. 

Figure \ref{figA1} shows that the different hybrid SFR methods are able to bring the FUV-SFRs of local galaxies into rough agreement with our best-fit MS and remove most of the inclination dependence. However, different calibrations can change the overall SFR$_\text{FUV,corr}$ by up to 0.2 dex. Table \ref{tabB1} gives the best-fit values for the different hybrid correction methods discussed above. WISE band 4 is less sensitive than W3, so we have about half the number of galaxies in our samples when we use 22$\mu$m corrections. Moreover, signal-to-noise ratio cuts were applied on SFRs and therefore depend on L$_{TIR}$, explaining the different number of galaxies for our analysis using W08.

\begin{table}
\caption{Local SFR$_\text{FUV,corr}$ from the hybrid relations indicated in Figure \ref{appendixfig}}\label{tabB1}
\begin{tabular}{lcccc}
\hline
Method & N & $k_1$ & $k_2$ & $\sigma^2$ \\
\hline 
12$\mu$m, C17 & 3453 & -0.15$^{+0.05}_{-0.05}$ & 0.315$^{+0.03}_{-0.03}$ & 0.174$^{+0.005}_{-0.005}$\\
22$\mu$m, C17 & 2111 & -0.15$^{+0.02}_{-0.02}$ & 0.32$^{+0.03}_{-0.03}$ & 0.20$^{+0.01}_{-0.01}$\\
12$\mu$m, W08 & 3453 & -0.15$^{+0.02}_{-0.02}$ & 0.15$^{+0.02}_{-0.02}$ & 0.190$^{+0.005}_{-0.005}$\\
22$\mu$m, W08 & 1562 & -0.14$^{+0.03}_{-0.03}$ & 0.10$^{+0.05}_{-0.05}$ &0.18 $^{+0.01}_{-0.01}$\\
22$\mu$m, CT15& 2111 & -0.15$^{+0.03}_{-0.03}$ & 0.10$^{+0.03}_{-0.03}$ & 0.22$^{+0.01}_{-0.01}$ \\

\hline
12$\mu$m, C17. & 1015 & -0.21$^{+0.05}_{-0.05}$ & 0.31$^{+0.02}_{-0.02}$ & 0.15$^{+0.01}_{-0.01}$ \\ 
22$\mu$m, C17. & 518 & -0.21$^{+0.08}_{-0.06}$ & 0.29$^{+0.03}_{-0.03}$ & 0.14$^{+0.01}_{-0.01}$\\
12$\mu$m, W08. & 1015& -0.21$^{+0.05}_{-0.05}$ & 0.14$^{+0.02}_{-0.02}$ & 0.16$^{+0.01}_{-0.01}$\\
22$\mu$m, W08. & 333 & 0.18$^{+0.10}_{-0.10}$ & 0.05$^{+0.05}_{-0.05}$ & 
0.12$^{+0.01}_{-0.01}$\\
22$\mu$m, CT15 & 518 &-0.21$^{+0.08}_{-0.08}$ & 0.06$^{+0.03}_{-0.03}$ & 0.165$^{+0.014}_{-0.014}$\\
\hline
\end{tabular}
\end{table}

\clearpage
\end{document}